\documentclass[prd,twocolumn,showpacs,preprintnumbers,floats,floatfix,
amssymb,amsmath,nofootinbib]{revtex4}
\usepackage{graphicx}
\usepackage{dcolumn}
\usepackage{latexsym}
\usepackage[latin1]{inputenc}
\usepackage{bm}
\begin{document}
\title{\Large \bf Soft Supersymmetry  Breaking on the Brane}
\author{Ph.  Brax}
\email{brax@spht.saclay.cea.fr}
\author{ N. Chatillon}
\email{chatillon@spht.saclay.cea.fr}
\affiliation{Service de Physique Th\'eorique, CEA/DSM/SPhT
Unit\'e de recherche associ\'ee au CNRS
CEA/Saclay
91191 Gif-sur-Yvette c\'edex, France
}
\vskip 1 cm

\begin{abstract}
We consider the low energy description of five dimensional models
of supergravity with boundaries comprising  a vector multiplet and
the universal hypermultiplet in the bulk. We analyse the
spontaneous breaking of supersymmetry induced by the vacuum
expectation value of superpotentials on the boundary branes. When
supersymmetry is broken, the moduli corresponding to the radion,
the zero mode of the vector multiplet scalar field and the dilaton
develop a potential in the effective action. We compute the
resulting soft breaking terms and give some indications on the
features of the corresponding  particle spectrum. We consider some
of the possible phenomenological implications when supersymmetry
is broken on the hidden brane.

\end{abstract}

\vskip 1 cm

\maketitle



\section{Introduction}
Supersymmetry breaking is one of the unsolved challenges of
particle physics. A proper understanding of the origin of
supersymmetry breaking would certainly increase the prospects of
an experimental discovery of supersymmetry and shed new light on
thorny issues such  as the cosmological constant problem. Many
models of supersymmetry breaking have been proposed so far.
Amongst the most popular are the gravity mediated and gauge
mediated scenarios (see \cite{gravity_gauge_mediated} for
reviews). Each have interesting features although none provide a
completely satisfactory framework. Recently brane models have been
introduced and address both the hierarchy problem \cite{ADD,RS1}
and the cosmological constant problem \cite{selftuning}. Brane
models have been originally built up in a non-supersymmetric
setting. The supersymmetrisation of the Randall-Sundrum model
\cite{Altendorfer:2000rr,Falkowski:2000er}
and models with a bulk scalar field \cite{kallosh,offshell,flp2} provide a
justification for certain fine--tunings used in brane models in
order to find flat brane solutions. Later it has been noticed that
supersymmetry is in fact compatible with branes of lower tension
than the Randall-Sundrum case
\cite{Brax:2001xf,susyRS_arbitrary_tensions,Brax:2002vs,
Lalak:2002kx,twisted_sugra,twist_warp_sugra}. When the tensions
are detuned, the 5d action of the theory is supersymmetric, but can
be built so that the warped background
solutions of the equations of motion either break supersymmetry
or not. Hence this may realize a
spontaneous breaking of supersymmetry in five dimensions. Other types
of supersymmetry breaking solutions with tension detuning, which do not
correspond to static warped backgrounds with straight branes, can lead to
strongly 4d Lorentz violating effects \cite{kinetic_susy_breaking,detuned}.

At sufficiently low energy, well below the brane tensions,
supersymmetric brane models can be described by a 4d
supersymmetric effective action
\cite{luty1,susy_radion,flp3,effective_detuned,effective_sugra}.
This 4d effective action is determined by a K\"ahler potential
 for the moduli fields. When the bulk contains a vector multiplet and the universal hypermultiplet, the moduli
describe the radion, the zero mode of the vector multiplet scalar
field and the dilaton, i.e. the zero mode of the hypermultiplet.
The moduli coming from the vector multiplet bulk scalars are
associated to the axion--like fields originating from the fifth
components of the bulk vector fields. The existence of these
massless moduli may imply strong deviations from general
relativity \cite{testsGR}. The cosmology of these tensor--scalar
theories is also interesting \cite{rev1,rev2}
 and may be probed
using the CMB anisotropies \cite{cosmoduli,CMB}.

The supersymmetric low energy action in the Randall-Sundrum case
has been nicely spelt out in \cite{effective_detuned}. Detuning
the brane tensions by including constant superpotentials leads to
an effective potential for the radion. It corresponds to the fact
that the detuned boundary conditions are no longer compatible with
4d Poincar\'e supersymmetry . The brane system is then subject to
a back-reaction effect which can be analysed using the equations
of motion of the low-energy action. When the radion effective
potential admits a minimum, this value of the radion is equal to
the one obtained by solving the 5d equations with detuned
tensions. At the minimum, the potential is negative and
supersymmetry is preserved for specific values of the radion
imaginary part, as the F--terms of the radion vanishes,
corresponding to $AdS_4$ supergravity. This is the same result as
obtained analysing the 5d equations of motion and Killing spinor
equations. Notice that the resulting configuration breaks
Poincar\'e invariance.

In the present article, we study the low energy action of brane
models with bulk scalar fields belonging to a vector multiplet and
the universal hypermultiplet of 5d $N=2$ supergravity. We include
the effects of a detuning of the brane tensions in the form of
superpotentials on the branes. Supersymmetry is then broken by the
$F$--terms associated to the moduli. We
analyse both the cosmological and the particle physics
consequences of such a breaking.

Our analysis of the breaking of supersymmetry extends to
the case with a vector multiplet the results of \cite{flp3} where
the Randall--Sundrum model with a hypermultiplet in the bulk was
considered. The phenomenology of this model has been spelt out in
\cite{casas} with particular emphasis on the electro--weak
symmetry breaking, for matter on the negative tension brane and
specific assumptions about moduli stabilisation. On the contrary, we
consider the case where matter lives on the positive tension brane with
no assumption about moduli stabilisation. We envisage the case when
moduli may not be stabilised and take into account the corresponding
solar system constraints.

A particular stumbling block of supersymmetric models is the
origin of the hierarchy between a large scale such as the Planck
mass and the $\mu$ term. In the following we will show that when
a $\mu$ term is included on the positive tension brane, a large
hierarchy can be induced thanks to the presence of the vector
multiplet scalar field in the bulk. Similarly, when supersymmetry
is broken on the hidden brane of negative tension, and no
$\mu$ term is included in the superpotential of the positive
tension brane, an effective $\mu$ term results from anomaly
mediation.

Breaking supersymmetry using boundary superpotentials has been
investigated in \cite{luty2,zwirner} in the flat case where the
brane tensions vanish. In the flat case, a $N=1$ superspace
formulation of $D=5$ supergravity
coupled to boundary branes has been elaborated in \cite{phil1} and used to compute
quantum corrections to the soft breaking terms \cite{phil2}. In the case that we consider, the bulk
background is warped. We analyse the non--trivial effects induced
by such a warping on the breaking of supersymmetry. In particular
all our results apply to the Randall--Sundrum setting where matter
lives on the positive tension brane.

When supersymmetry is broken on the hidden brane of negative
tension and matter lives on the positive tension brane, we find
that soft supersymmetry breaking terms are of two sorts. First of
all the soft trilinear $A$ terms and the gaugino masses receive a
non--vanishing contribution at one--loop level from  anomaly
mediation. Secondly  and contrarily to anomaly mediation
scenarios,  the soft masses are non-vanishing at tree level.
Therefore they do not suffer from the tachyons  of anomaly
mediated models.

In section 2 we present the models including a bulk vector
multiplet and study the zero--modes, i.e. we give different
parametrisations of the low energy moduli.  In section 3 we
analyse the low energy action computing the K\"ahler potential and
the superpotential when coupling 5d supergravity with a vector
multiplet in the bulk to matter on the branes. We also discuss the
dilaton arising from the bulk hypermultiplet.  In section 4 we
focus on the case with no hypermultiplet and compute both the
moduli potential and the soft breaking terms at the classical
level. We then introduce the dilaton field in section 5 and
discuss race--track models. In section 6, we discuss the
one--loop anomaly mediated soft terms. In section 7, we focus  on
supersymmetry breaking on the hidden brane, making explicit the
soft breaking terms and the associated phenomenology. In section
8, we introduce an explicitly supersymmetry breaking step by
taking into account  charges on the branes in order to bypass the
cosmological constant problem. This leads to an extra contribution
to the moduli potential. Finally in section 9, we discuss the
cosmological consequences and the gravitational constraints on the
models. We also include two appendices. In a first appendix, we discuss
the Randall-Sundrum case and radion stabilisation. In the second one,
we give the Randall-Sundrum soft terms.

\section{Supergravity with Boundary Branes}
\subsection{Supergravity Construction}
For the bulk theory with no brane coupling, ${\mathcal N}=2$ $D=5$
pure supergravity was first constructed in \cite{5d pure sugra}~;
vector multiplets were added in \cite{vector coupling}, and
finally vector multiplets, hypermultiplets and tensor multiplets
were treated together in \cite{vector-tensor-hyper}. Gauged
supergravity with boundary branes in five dimensions has been
elegantly constructed when vector multiplets live in the bulk
\cite{kallosh,offshell}. The supergravity multiplet comprises the
metric tensor $g_{ab}$, $a,b=1\dots 5$, the gravitini $\psi_a^A$
where $A=1,2$ is an $SU(2)_R$ index and the graviphoton field
$A_a$. The ${\mathcal N}$=2 vector multiplets in the bulk possess
one vector field, a $SU(2)_R$ doublet of symplectic Majorana
spinors and one real scalar. When considering $n$ vectors
multiplets, it is convenient to denote by $A^I_a$, $ I=1\dots
n+1$, the $(n+1)$ vector fields.

The two boundaries are fixed points of a $Z_2$ orbifold like in the Randall-Sundrum model
and its supersymmetrisation. The action on each brane depends on two ingredients. First the branes couple
to the bulk, i.e. to gravity and the real scalar fields. Then ordinary matter is confined to either of the branes.
In the following, we will first describe the bulk and brane theory without matter.

The  supergravity theory with boundaries differs from usual
five-dimensional non supersymmetric theories with boundaries as
new superpartner fields are introduced in order to close the
supersymmetry algebra and ensure the invariance of the Lagrangian.
The vector multiplets comprise scalar fields $\phi^i$
parameterizing the manifold
\begin{equation}
C_{IJK}h^I(\phi) h^J(\phi) h^K(\phi)=1
\end{equation}
with the functions $h^I(\phi), I=1\dots n+1$ playing the role of
auxiliary variables. In heterotic M--theory \cite{ovrut} the
symmetric tensor $C_{IJK}$ has the meaning of an intersection
tensor. Defining the metric
\begin{equation}
G_{IJ}=-2C_{IJK}h^{K}+3h_Ih_J
\end{equation}
where $h_I=C_{IJK}h^Jh^K$, the bosonic part of the Lagrangian (vector fields
not included) reads
\begin{equation}
S_{bulk}=\frac{1}{2\kappa_5^2}\int d^5x \sqrt {-g_5}\Big({\mathcal
R} -\frac{3}{4}(g_{ij}\partial_{\mu}
\phi^i\partial^{\mu}\phi^j+V)\Big) \label{lag}
\end{equation}
where the sigma-model metric $g_{ij}$ is
\begin{equation}
g_{ij}=2 G_{IJ}\frac{\partial h^I}{\partial \phi^i}\frac{\partial h^J}{\partial \phi^j}
\end{equation}
and the potential is given by
\begin{equation}
V=U_iU^i-U^2
\end{equation}
using the sigma-model metric $g_{ij}$. The superpotential $U$ defines the dynamics of the theory.
It is given  by
\begin{equation}
U=4\sqrt\frac{2}{3}g h^IV_I
\end{equation}
where $g$ is a gauge coupling constant and the $V_I$'s are real
numbers such that the $U(1)$ gauge field is $A^I_a V_I$.

The boundary action depends on two new fields. There is a
supersymmetry singlet  $G$ and a four form $A_{\mu\nu\rho\sigma}$
\cite{kallosh}. One also modifies  the bulk action by replacing
$g\to G$ and adding  a direct coupling
\begin{equation}
S_A=\frac{2}{4! \kappa_5^2}\int d^5x \epsilon^{abcde}A_{abcd}\partial_{e}G.
\end{equation}
The boundary action is taken as
\begin{widetext}
\begin{equation}
S_{bound}=-\frac{1}{\kappa_5^2}\int d^5x
(\delta_{x_5}-\delta_{x_5-R}) (\sqrt {-g_4}\frac{3}{2}U
+\frac{2g}{4!}\epsilon^{\mu\nu\rho\sigma}A_{\mu\nu\rho\sigma}).
\label{boundary action}
\end{equation}
\end{widetext}
where $\mu,\nu,\rho,\sigma$ are four-dimensional indices on the
branes. Notice that the four-form $A_{abcd}$ is not dynamical.
According to this action the branes can be seen as charged under
this bulk four-form with a charge $\pm g$.  In section \ref{charged branes} we will
consider branes charged under a new bulk four-form with kinetic
terms and arbitrary charge, in order to cancel the vacuum energy.

The supersymmetry algebra closes on shell where
\begin{equation}
G(x)=g\epsilon (x_5),
\end{equation}
and $\epsilon (x_5)$ jumps from -1 to 1 at the origin of the fifth dimension.
On shell the bosonic Lagrangian reduces to  the bulk Lagrangian  coupled to the
boundaries as,
\begin{equation}
S_{bound}=-\frac{3}{2\kappa_5^2}\int d^5x
(\delta_{x_5}-\delta_{x_5-R})\sqrt {-g_4}U,
\end{equation}
Crucially, the boundary branes couple directly to the bulk
superpotential. Notice that the two branes have opposite
(field-dependent) tensions
\begin{equation}
\lambda_{\pm}=\pm \frac{3}{2\kappa_5^2}U
\end{equation}
where the first brane has positive tension.

Let us focus on the case of a single vector multiplet $n=1$. The
equations of motion can be written in a first order BPS form
\begin{equation}
\frac{a'}{a}=-\frac{U}{4},\ \phi'=\frac{\partial U}{\partial \phi},
\end{equation}
where $'=d/dz$ for a metric of the form
\begin{equation}\label{background}
ds^2 = dz^2 + a^2(z)\eta_{\mu\nu}dx^\mu dx^\nu.
\end{equation}
The boundary conditions are automatically satisfied implying that
the positions of the two boundary branes are not specified.
Moreover the BPS background preserves $N=1$ Poincar\'e supersymmetry from the
four dimensional point of view. Indeed denote by $\epsilon^A$ the
supersymmetry parameter of $N=2$ 5d supergravity. The Killing
spinor solutions of $\delta_\epsilon \psi_a^A=0$
satisfy \cite{kallosh}
\begin{equation}
\epsilon ^A (z,x^\mu) =a^{1/2}(z) \epsilon^A
\end{equation}
where $\epsilon^A$ is a constant spinor such that  $\gamma_5
\epsilon^A =(\sigma_3)^A_B\epsilon^B$ implying that only one
chirality of the original supersymmetries is preserved. Having
obtained Killing spinors corresponding to $N=1$ 4d supersymmetry,
we will explicitly find that the low energy Lagrangian can be
written in a 4d supersymmetric way.

Let us give the simplest example of models of supergravity with a
single scalar field \cite{BD}. We choose only one vector multiplet
and the only component for  the symmetric tensor $C_{IJK}$ is
$C_{122}=1$. The moduli space of vector multiplets is then defined
by the algebraic relation
\begin{equation}
3 h^1(h^2)^2=1.
\end{equation}
This allows to parameterize this  manifold using the coordinate
$\phi$ such that
 $h^1$ is proportional to $e^{\sqrt{\frac{1}{3}}\phi}$
and $h^2$ to $e^{-\phi/2\sqrt 3}$. The induced metric $g_{\phi\phi}$ can be
seen  to be one. The most general superpotential is a linear combination of the two exponentials
$U=ae^{\sqrt{\frac{1}{3}}\phi}+be^{-\phi/2\sqrt 3}$.
In the following we will focus on models where the superpotential $U$ can be expressed as an exponential
of the normalised scalar field $\phi$
\begin{equation}\label{potential}
U=4k e^{\alpha \phi}.
\end{equation}
The values  $\alpha =1/\sqrt 3,-1/\sqrt {12}$ correspond to the
previous example. The metric in the bulk depends on the scale
factor
\begin{equation}\label{scale}
a(z)=(1-4k\alpha^2z)^{\frac{1}{4\alpha^2}},
\end{equation}
while the scalar field solution is
\begin{equation}\label{psi}
\phi = -\frac{1}{\alpha}\ln\left(1-4k\alpha^2z\right).
\end{equation}
In the $\alpha\to 0$ we retrieve the AdS profile
\begin{equation}
a(z)=e^{-kz}.
\end{equation}
corresponding to the supersymmetric Randall-Sundrum model with no vector multiplet in the bulk.
\subsection{The zero--modes}

The BPS configurations have zero modes solving the linearised
equations of motion together with the boundary conditions at the
branes. The linearised Einstein and Klein--Gordon equations have
been thoroughly studied in \cite{chris1} following an earlier work
\cite{christos} in the Randall--Sundrum case. The end result is
that there are two scalar modes and one spin two mode. They
correspond to either the radion and the zero mode of the bulk
scalar field or the two brane positions. The spin two zero mode,
i.e. the graviton, is associated to 4d gravity at low energy.

The number of moduli can be inferred by a counting argument based
on supersymmetry. At low energy, the only imprint of the two bulk
vector fields $A_a^I,\ I=1,2,$ are two pseudo--scalar fields
$A_5^I$ (the vector fields are projected out by  the $Z_2$
symmetry). These two axion--like fields are combined with the
radion and the bulk scalar field to form the scalar fields
comprising two chiral superfields. Already in the Randall--Sundrum
case, the radion can be seen as the fluctuation of the $g_{55}$
component of the bulk metric and is associated to the zero mode of
the gravi--photon $A_5$. This remains true and is complemented by
the association of the bulk scalar field zero mode of the vector
multiplet to the corresponding axion--like field in the vector
multiplet.

The two scalar zero modes can be viewed as the two brane positions
\begin{equation}
t_1=\xi_1(x),\ t_2= r+\xi_2(x)
\end{equation}
representing the massless fluctuations with respect to fixed
branes at $0$ and $r$
\begin{equation}
\Box^{(4)}\xi_{1,2}=0
\end{equation}
where the bulk metric is unperturbed.

Equivalently the two branes can be considered as fixed and the
metric is perturbed
\begin{equation}
ds^2= a^2(G(x,z))g_{\mu\nu}dx^\mu dx^\nu + (\partial_z G)^2 dz^2
\end{equation}
where $g_{\mu\nu}= \eta_{\mu\nu} +h_{\mu\nu}$ is the perturbed 4d
metric and
\begin{equation}
G(x,z)= z +\frac{\xi(x)}{a^2(z)} +\xi_0(x)
\end{equation}
where
\begin{equation}
\Box^{(4)} \xi= 0, \ \Box^{(4)} \xi_0= 0
\end{equation}
and
\begin{equation}
\xi_1=\frac{\xi}{a_1^2}+\xi_0,\ \xi_2=\frac{\xi}{a_2^2}+\xi_0
\end{equation}
The radion is related to $\xi$ as
\begin{equation}
t(x)=(\frac{1}{a_2^2}-\frac{1}{a_1^2})\xi(x)
\end{equation}
where $a_{1,2}$ are the scale factors of the first and second
branes. The mode $\xi_0$ is associated  to a global translation of
the bulk--brane system. Similarly the scalar field is perturbed as
\begin{equation}
\phi(x,z)=\phi_{BPS}(G(x,z))
\end{equation}
No intrinsic zero--mode is associated to the scalar field.

To linear order, this parametrisation is equivalent to
\begin{equation}
\delta g_{55}= -2 \frac{a'}{a^3} \xi
\end{equation}
confirming the link between $\delta g_{55}$ and the radion.
Similarly the perturbed scalar field is
\begin{equation}
\frac{\delta \phi}{\phi'_{BPS}}= 2\frac{\delta g_{55}}{U}
+\frac{\partial U}{\partial \phi}\xi_0
\end{equation}
picking contributions from $\delta g_{55}$ and $\xi_0$.

Finally, one can also use a parametrisation generalising the one
of \cite{effective_detuned}
\begin{equation}
ds^2= A^2(x,z) \Omega^2 (x,z) g_{\mu\nu} dx^\mu dx^\nu + \frac{(
a^2+ 2aa' \xi_0)^2}{A^4}dz^2
\end{equation}
where
\begin{equation}
A^2= a^2 + 2\frac{a'}{a}\xi +2 aa' \xi_0
\end{equation}
and
\begin{equation}
\Omega^2= (1- \frac{a_1^2-a_2^2}{\int_0^r a^2 dz} \xi_0)^{-1}
\end{equation}
The dimensional reduction using this ansatz leads to an action in
the Einstein frame for the two moduli $\xi_0$ and $\xi$.

In the following, we will work exclusively with the two moduli
$t_{1,2}$ as the parametrisation is simpler.

\section{The Low Energy Action}

\subsection{The vector multiplet sector}

At low energy the brane and bulk system is amenable to a
four--dimensional treatment where the dynamics are captured by the
slow motion of moduli fields. Two of the moduli of the system are
the brane positions as they are not specified by the equations of
motions.   At low energy, one considers small deformations of the
static configuration allowing the moduli to be space--time
dependent.  We denote the position of brane 1 by $t_1(x^\mu)$ and
the position of brane 2 by  $t_2(x^\mu)$. We consider the case
where the evolution of the brane is slow. This means that in
constructing the effective four--dimensional theory we neglect
terms of order  higher than two in a  derivative expansion.
Moreover the non-linearity of the Einstein equations on the branes
in the matter energy--momentum tensors \cite{project} are
neglected. Such a regime is only valid at low energy well below
the brane tensions of both branes. For instance, putting the
standard model matter on the positive tension brane leads to an
effective action valid only up to the positive brane tension. In
addition to the brane positions, we need to include the graviton
zero mode, which can be done by replacing $\eta_{\mu\nu}$ with a
space--time dependent tensor $g_{\mu\nu}(x^\mu)$.

After integrating over the fifth dimension, one obtains a 4d
effective action. The Einstein-Hilbert term in 4d follows from the
5d term and reads
\begin{equation}
S_{\rm bulk} = \int d^4 x \sqrt{-g_4} f(t_1,t_2) {\cal R}^{(4)},
\end{equation}
with
\begin{equation}
f(t_1,t_2) = \frac{1}{\kappa_5^2} \int^{t_2}_{t_1} dz a^2 (z)
\end{equation}
For the exponential superpotential $U$, this is
\begin{equation}
f(t_1,t_2)=
\frac{a(t_1)^{2+4\alpha^2}-a(t_2)^{2+4\alpha^2}}{(2+4\alpha^2)k\kappa_5^2}.
\end{equation}
Notice that the action is in the brane  frame different from the
Einstein frame. Including the boundary terms  leads to the
following effective action \cite{cosmoduli}
\begin{widetext}
\begin{eqnarray}
S_{} = \int d^4 x \sqrt{-g_4}\left[ f(t_1,t_2) {\cal R}^{(4)} +
\frac{3}{4}a^2(t_1)\frac{U(t_1)}{\kappa_5^2}(\partial t_1)^2 -
\frac{3}{4} a^2(t_2)\frac{U(t_2)}{\kappa_5^2}(\partial t_2)^2
\right].
\end{eqnarray}
\end{widetext}
As expected the moduli are free scalar fields.

The effective action for the two moduli $t_1$ and $t_2$ is written
in an explicit supergravity form. This follows from the fact that
the  two-brane system satisfies BPS conditions. At low energy the
bulk and brane system preserves one of the original
supersymmetries. Indeed one can write the Einstein-Hilbert term
and the kinetic terms of the moduli as
\begin{widetext}
\begin{equation}
 \int d^4 x \sqrt{-g_4}\left[ f(t_1,t_2) {\cal R}^{(4)}
+ 6 \partial_{T_1}\partial_{\bar T_1}f\partial_\mu T_1\partial^\mu
\bar T_1 + 6\partial_{T_2}\partial_{\bar T_2}f\partial_\mu T_2
\partial^\mu \bar T_2\right ],
\end{equation}
\end{widetext}
where
\begin{equation}
t_1=\frac{1}{2}(T_1+\bar T_1),\ t_2=\frac{1}{2}(T_2+\bar T_2).
\end{equation}
This allows us to identify  the fields $T_1$ and $T_2$ as the
scalar parts of two chiral multiplets whose dynamics are captured
by the function $f(t_1,t_2)$
\begin{equation}
\int d^4 x d^4 \theta E^{-1} f(\frac{T_1+\bar T_1}{2}\
,\frac{T_2+\bar T_2}{2})
\end{equation}
where $E$ is the vielbein determinant superfield. From this action
one can read off the K\"ahler potential for the two moduli fields
in the Einstein frame
\begin{equation}
K=-3\ln (\kappa_4^2 f).
\end{equation}
which depends on the two moduli.
 Notice that the K\"ahler potential
possesses two global symmetries
\begin{equation}
T_i\to T_i +ib_i .
\end{equation}
coming from the independence of the background geometry on the
axion fields. A detailed analysis of the K\"ahler geometry with an
arbitrary number of vector multiplets is under study \cite{kahler}.

Let us now concentrate on the Randall--Sundrum case $\alpha=0$,
the K\"ahler potential reads
\begin{equation}
K=-3\ln ( e^{-k(T_1+\bar T_1)}-e^{-k(T_2+\bar T_2)}).
\end{equation}
The K\"ahler potential can be written as
\begin{equation}
K=3k(T_1+\bar T_1)-3\ln(1-e^{-k(T+\bar T)}),
\end{equation}
where
\begin{equation}
T=T_2- T_1
\end{equation}
is the radion superfield.  In the Randall--Sundrum case, the field
$T_1$ can be eliminated by a K\"ahler transformation; this shows
that one of the two moduli decouples, leaving only the radion as
the relevant physical field.

\subsection{The hypermultiplet sector}

We can now introduce another ingredient, i.e. a hypermultiplet
living in the bulk. We will focus on the universal hypermultiplet
comprising four scalar fields, two being odd under the orbifold
parity \cite{flp2}. We also assume that the hypermultiplet is not
charged under the gauged $U(1)_R$ symmetry in such a way that no
contribution from the hypermultiplet appears in the bulk
potential. At low energy, the two even scalar fields become the
scalar part of a chiral multiplet $S$ which will be called the
dilaton in the following. The low energy dynamics of the dilaton
is determined by the K\"ahler potential
\begin{equation}
K(S,\bar S)=- \ln (S+\bar S)
\end{equation}
Notice that the full moduli K\"ahler potential is then just the
sum of the vector multiplet and hypermultiplet contributions.

\subsection{The coupling to matter \label{section coupling}}

 Let us now introduce matter on
the boundary branes. We couple the matter fields to the induced
metric on the ith--brane leading to an action for the matter
scalar field $s$ coupled to the brane position $t_i$
\begin{equation}
\int d^4 x \sqrt{-g} \Big( a^2(t) (\partial s\partial \bar s)+ a^4(t)
\vert\frac{\partial w(s)}{\partial s}\vert ^2 \Big)
\end{equation}
up to derivative terms in $t_i$. We have denoted by $w$ the
superpotential of the supersymmetric theory on the brane. As
we are supersymmetrising the matter action only at zeroth order in $\kappa_4$, we have
suppressed the non-renormalizable terms in the matter fields for fixed moduli, hence the globally  supersymmetric  form of the potential. Such an
action can be supersymmetrised (we follow the conventions of \cite{weinberg} for the definitions of the K\"ahler potentials)
\begin{equation}
-3\int d^4 x d^4 \theta E^{-1} \Big( f-  a^2(\frac{T_i +\bar
T_i}{2})\Sigma \bar \Sigma\Big)
\end{equation}
modifying the K\"ahler potential of the moduli
\begin{equation}
K=-3\ln (\kappa_4^2 f - \frac{1}{3}\kappa_4^2 a^2(\frac{T_i+\bar
T_i}{2})\Sigma \bar \Sigma) \label{kha}
\end{equation}
where $\Sigma =s+\dots $ is the chiral superfield of matter on the
brane (not to be confused with the hypermultiplet superfield S).
Similarly the potential on the brane follows from
\begin{equation}
\int d^4x d^2\theta \Phi^3 W(T_i,S,\Sigma)
\end{equation}
where
\begin{equation}
W(T_i,S)= a^3(T_i)w( S,\Sigma)
\end{equation}
and $\Phi$ is the chiral compensator whose $F$--term is the
gravitational auxiliary field.  At low energy this leads to a
direct coupling between matter fields and the moduli. This is
crucial when discussing supersymmetry breaking. Moreover the
coupling to the brane breaks the global symmetries $T_i\to T_i +
ib_i$. This allows to break supersymmetry using the imaginary
parts (axions) of the moduli fields. In the Randall--Sundrum case,
the symmetry is extended to $SL(2,\mathbb{R})$ which is not broken
by the boundary superpotential \cite{effective_sugra}. This leads
to the absence of any axionic dependence in the scalar potential \cite{effective_detuned}.

When including matter fields on several branes, the superpotential
is simply a sum of all the contributions coming from the different
branes. The K\"ahler potential is obtained by summing the different
contributions from the branes inside the logarithmic term.
\\

Let us now turn to the gauge sectors. We assume that each brane
carries gauge fields $A_\mu^{1,2}$ associated to the gauge groups
$G_{1,2}$. As the gauge kinetic terms are conformally invariant we
find that the gauge coupling constant does not acquire a moduli
dependence. However, we need a non constant gauge coupling to
generate gaugino masses. For example, one can make it explicitly
dependent on the brane value of the bulk scalar field $\phi(T_i)=
-4\alpha \ln(a(T_i))$ in the 5d action. Contrarily to the brane
localized potential of the bulk scalar field (the tension), which
is related to its bulk potential, this dependence is not
constrained by local brane-bulk supersymmetry and is thus
arbitrary. In that case however, by 5d locality the gauge coupling
function on brane $i$ can only depend (analytically) on the $T_i$
modulus (and the dilaton S). In the absence of more information on
the coupling of the bulk scalar field to the brane gauge kinetic
terms, we will allow for a general coupling to the moduli fields
\begin{equation}
\int d^4 x d^2 \theta (f_1(T_1,S) {\mathcal W}_{1\alpha} {\mathcal
W}^\alpha_1 +f_2(T_2,S) {\mathcal W}_{2\alpha} {\mathcal
W}^\alpha_2)
\end{equation}
Another possibility is to consider the anomalous breaking of the
(super-)conformal invariance of the gauge kinetic term \cite{anomaly}. First,
conformal invariance of the action before gauge fixing the
conformal compensator superfield $\Phi$ to $1+\theta^2 F_{\Phi}$
implies that $\Phi$ must multiply any cut-off dependence appearing
during the renormalisation process. Furthermore, after the
gauge-fixing of $\Phi$, a R-symmetry transformation of the action
results in an anomalous shift of the gauge $\theta$-angle given by
the imaginary part of the gauge coupling. Compensation of this
shift by a R-symmetry phase rotation of $\Phi$ before gauge-fixing
dictates the $\Phi$ dependence for the gauge coupling function
$f_a$ evaluated at the scale $E$ on either of the branes
\begin{equation}
f_a(E)=f_a^0 + \frac{b}{2\pi} \ln(\frac{E}{\Lambda_{UV}(T_1,T_2)
\Phi})
\end{equation}
where the effective  cut-off $\Lambda_{UV}$ depends on the moduli
$T_i$ in the Einstein frame. Anomaly mediation  contributes to the
gaugino masses both through $F_{\Phi}$ and $F^{T_i}$. We present
this possibility in section \ref{AMSB}.

\section{ Supersymmetry Breaking at Tree Level}
\subsection{F--type supersymmetry breaking}

We will consider the case of a single vector multiplet in the bulk
and its coupling to the two boundary branes carrying  matter
superfields $\Sigma_{1,2}^i$. The case including the universal hypermultiplet will be dealt with later. The low energy theory depends on the
K\"ahler potential and the superpotential. The K\"ahler potential
includes both the matter fields and moduli
\begin{widetext}
\begin{eqnarray}
K &=& -3\ln \Big[\kappa_4^2 f(\frac{T_1+ \bar
T_1}{2},\frac{T_2+\bar T_2}{2}) - \frac{1}{3}\kappa_4^2 a^2(\frac{T_1+\bar
T_1}{2}) (\Sigma_1^i \bar \Sigma_1^{\bar i}+
\frac{\lambda^1_{ij}\Sigma_1^i\Sigma_1^j+ cc}{2})
\nonumber \\
&& - \frac{1}{3}\kappa_4^2 a^2(\frac{T_2+\bar T_2}{2})(\Sigma_2^i \bar
\Sigma_2^{\bar i}+ \frac{\lambda^2_{ij}\Sigma_2^i\Sigma_2^j+
cc}{2})\Big] \label{kha_with_GM}
\end{eqnarray}
\end{widetext}
We consider diagonal kinetic terms and Giudice-Masiero mixing
terms \cite{GM} $\lambda^{1,2}_{ij}$ for phenomenological purpose,
i.e in order to address the so--called $\mu$ problem. We expand
the superpotential
\begin{equation}
W=a(T_1)^3 w(\Sigma_1^i)+a(T_2)^3w(\Sigma_2^i)
\end{equation}
Notice that the two sectors on the branes are decoupled in the
superpotential. The matter superpotentials are
\begin{equation}
w(\Sigma_{1,2}^i)= w_{1,2} + \frac{1}{2}\mu_{ij}^{1,2} \Sigma_{1,2}^i \Sigma_{1,2}^j
+\frac{1}{6}\lambda_{ijk}^{1,2} \Sigma_{1,2}^i \Sigma_{1,2}^j\Sigma_{1,2}^k
\end{equation}
where the constant pieces $w_{1,2}$ give a negative contribution
to the brane cosmological constants, i.e. the brane tensions.
Incorporating  constant terms in the superpotentials of each
brane, one obtains a brane configuration with detuned tensions.
The detuning of the brane tensions is responsible for
supersymmetry breaking when the F--terms of $T_1$ or $T_2$ are non-vanishing.

Using the previous ingredients one can work out  the soft
supersymmetry breaking terms in the Einstein frame. In the
following we will focus on the exponential coupling $U=4k
e^{\alpha \phi}$. We identify the inverse squared Planck mass as
\begin{equation}
\kappa_4^2= 2k (1+2\alpha^2) \kappa_5^2
\end{equation}
The gravitino mass is given by $m_{3/2}=\kappa_4^2 <e^{K/2}\vert
W\vert >$ where the non-vanishing vev is provided by the constant
terms in the superpotentials. Moreover, the gravitino mass becomes a function
of the moduli
\begin{equation}
m_{3/2}= \kappa_4^2 \Delta^{-3/2} \vert a(T_1)^3 w_1 +a(T_2)^3
w_2\vert
\end{equation}
where $\Delta=a(t_1)^{2+4\alpha^2}-a(t_2)^{2+4\alpha^2}$.
 In the Randall-Sundrum case with $\alpha=0$, this reduces
to
\begin{equation}
m_{3/2}= \kappa_4^2 \frac{\vert w_1 +e^{-3kT}
w_2\vert}{(1-e^{-2kt})^{3/2}}
\end{equation}
It is interesting to compute the F--terms associated to
the breaking of supersymmetry. The non--vanishing F--terms
associated to the two moduli are
\begin{widetext}
\begin{eqnarray}
F^{T_1}&=& <\frac{W}{|W|}>\Big[2 \kappa_5^2\Delta^{-1/2}a(\bar
T_1)^{3-4\alpha^2} a(t_1)^{-2+4\alpha^2}\bar w_1
\nonumber \\
&&+ 4 (1+2\alpha^2)\Delta^{-3/2}
k \kappa_5^2 \ i\ {\cal I}m T_1\ a(\bar T_1)^{3-4\alpha^2}
a(t_1)^{4\alpha^2}\bar w_1
\nonumber \\
&&+ 4 (1+2\alpha^2)\Delta^{-3/2} k \kappa_5^2 \ i\ {\cal I}m T_2\
a(\bar T_2)^{3-4\alpha^2} a(t_1)^{4\alpha^2}\bar w_2\Big]
\nonumber \\
F^{T_2}&=& <\frac{W}{|W|}>\Big[-2 \kappa_5^2\Delta^{-1/2}a(\bar
T_2)^{3-4\alpha^2} a(t_2)^{-2+4\alpha^2}\bar w_2
\nonumber \\
&&+ 4 (1+2\alpha^2)\Delta^{-3/2}
k \kappa_5^2 \ i\ {\cal I}m T_2\ a(\bar T_2)^{3-4\alpha^2}
a(t_2)^{4\alpha^2}\bar w_2
\nonumber \\
&&+ 4 (1+2\alpha^2)\Delta^{-3/2} k \kappa_5^2 \ i\ {\cal I}m T_1\
a(\bar T_1)^{3-4\alpha^2} a(t_2)^{4\alpha^2}\bar w_1\Big]
\nonumber \\
\end{eqnarray}
\end{widetext}
where we have defined F-terms using the following phase convention
:
\begin{equation}
F^i \equiv e^{K/2}|W| G^{i \bar j}G_{\bar j} =
\frac{W}{|W|}e^{K/2}K^{i \bar j}(\overline{W}_{\bar j} +
\kappa_4^2 \overline{W} K_{\bar j}).
\end{equation}
As the moduli are of length  dimension one, the F--terms are
dimension--less. In the case of vanishing imaginary parts of the
moduli $T_i$, they simplify to
\begin{eqnarray}
F^{T_1}&=& 2 \kappa_5^2\Delta^{-1/2}a(t_1) \bar w_1  <\frac{W}{|W|}> \nonumber \\
F^{T_2}&=& -2 \kappa_5^2\Delta^{-1/2} a(t_2) \bar w_2 <\frac{W}{|W|}> \nonumber \\
\end{eqnarray}
Notice that in this case, each modulus breaks supersymmetry when
the constant superpotential on the corresponding brane is
non--zero. The soft supersymmetry breaking terms are a direct
consequence of the detuning of the brane tensions. Note that the
dependence of supersymmetry breaking on the vev of the radion
imaginary part has also been studied in \cite{wilson} in the case
of the detuned Randall-Sundrum model.

\subsection{The moduli potential \label{potential section}}

The breaking of supersymmetry leads to a potential for the moduli
fields given by $V=\kappa_4^{-4}e^{G}(G_iG^i -3)$ where $G= K +\ln
\kappa_4^6 \vert W\vert^2$. This gives the potential
\begin{widetext}
\begin{eqnarray}
V&=&\frac{3\kappa_4^2}{1+2\alpha^2}\Delta^{-2}\Big[ \vert
a(T_2)^{3-4\alpha^2} w_2\vert^2 a(t_2)^{-2+4\alpha^2}-\vert
a(T_1)^{3-4\alpha^2} w_1\vert^2 a(t_1)^{-2+4\alpha^2}\Big]
\nonumber \\
&&+24\alpha^2 k^2 \kappa_4^2 \Delta^{-3}\vert a(T_1)^{3-4\alpha^2}
w_1 {\cal I}m T_1 +a(T_2)^{3-4\alpha^2} w_2 {\cal I}m T_2\vert^2 .
\label{pot with im non zero}
\end{eqnarray}
\end{widetext}
We have distinguished the real terms $a(t_{1,2})$ from the complex
terms $a(T_{1,2})$ which depend on the axion--like fields. Notice
that as soon as one of the $w_i$ is vanishing, the
corresponding ${\cal I}m T_i$ becomes a flat direction in
agreement with the restoration of the global symmetry $T_i \to T_i
+ib_i$.

Let us
first consider the Randall-Sundrum case $\alpha=0$. The potential
does not depend on the axion--like fields at all
\begin{equation}
V= {3\kappa_4^2}\frac{|w_2|^2 \rho^{4}-|w_1|^2}{(1-\rho^{2})^2} .
\end{equation}
We have  defined $\rho =a(t_2)/a(t_1)$. Hence the axion--like
field ${\cal I}m T$ is a flat direction  while the radion  flat direction is
lifted.

Notice that for $\alpha \ne 0$ the flat directions for the
axion--like fields ${\cal I} m T_i$ are lifted. One can show that
\begin{equation}
({\cal I} m T_1,{\cal I} m T_2)=(0,0)
\end{equation}
is an extremum of the potential. In the scenario of hidden brane
supersymmetry breaking ($w_1=0$, $w_2\neq 0$) that we will
consider later, this extremum is stable in ${\cal I} m T_2$, while
${\cal I} m T_1$ becomes a flat direction of the potential as
already mentioned.

The potential becomes then
\begin{equation}
V=
\frac{3\kappa_4^2}{1+2\alpha^2}a(t_1)^{-12\alpha^2}\frac{|w_2|^2
\rho^{4-4\alpha^2}-|w_1|^2}{(1-\rho^{2+4\alpha^2})^2} .
\label{pot}
\end{equation}
This is equivalent to the potential obtained by detuning the brane
tensions
\begin{equation}
\lambda_{\pm}=\pm \delta _{1,2} \frac{3}{2\kappa_5^2} U .
\end{equation}
where the detuning parameter $\delta_{1,2}$ is less than one
\begin{equation}
\delta_{1,2}= 1-\kappa_5^2 |w_{1,2}|^2 .
\end{equation}
Notice that the brane tension in  supergravity
 is always less than the tuned tension with no
supersymmetry breaking \cite{Brax:2001xf,susyRS_arbitrary_tensions}.

  Several cases may
be distinguished, which are summarized in figures \ref{figure
potentiels 1}, \ref{figure potentiels 3} and \ref{figure
potentiels 6} for $Q=0$. Note that only one half of the different
profiles is actually possible for a vanishing $Q$ parameter (whose
significance as a brane charge will be explained in section
\ref{charged branes}). When $w_1\ne 0$ and $|w_2|\leq |w_1|$ , the
theory is unstable, with an unbounded from below potential in the
limit $\rho \to 1$, i.e. for nearly colliding branes. This
corresponds to figure \ref{figure potentiels 3}. For
$w_1\ne 0$ and $|w_2|>|w_1|$, the potential has a global minimum
but at a negative energy ; this is figure \ref{figure potentiels
1}. Finally, the case $w_1=0$ is phenomenologically
more interesting as the potential is positive and bounded from
below ; this is figure \ref{figure potentiels 6}. All the
other figures correspond to $Q\ne 0$ and will be considered later
in section \ref{charged branes}.


\begin{figure}[!h]
\begin{center}
\includegraphics[height=8cm]{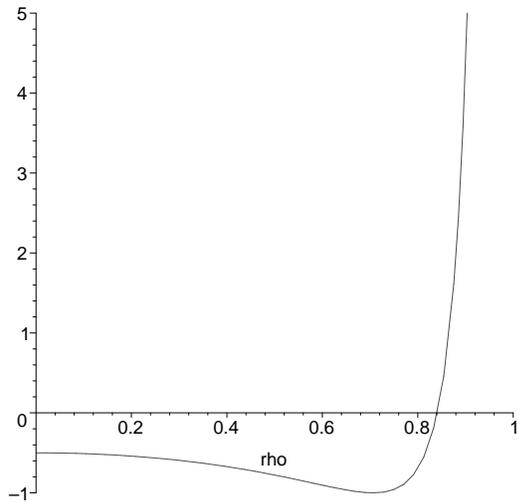}
\end{center}
\caption{The moduli potential as a function of
$\rho=a(t_2)/a(t_1)$ for ${\mathcal I}m T_{1,2}=0$
and $ \vert w_2\vert > \vert w_1 \vert > \vert Q\vert$. The radion
may be stabilised with a negative vacuum energy.}
\label{figure potentiels 1}
\end{figure}

\begin{figure}[!h]
\begin{center}
\includegraphics[height=8cm]{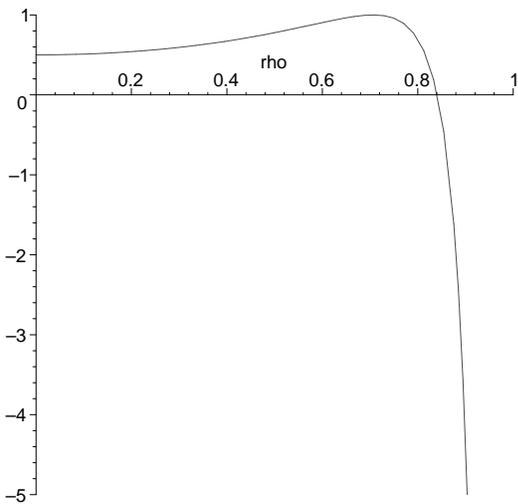}
\end{center}
\caption{The moduli potential as a function of
$\rho=a(t_2)/a(t_1)$ for ${\mathcal I}m T_{1,2}=0$ and
$ \vert w_2\vert < \vert w_1 \vert < \vert Q\vert$. The radion
has a metastable minimum for an infinite interbrane distance, and
an unbounded from below branch leading to colliding branes.}
\label{figure potentiels 2}
\end{figure}

\begin{figure}[!h]
\begin{center}
\includegraphics[height=8cm]{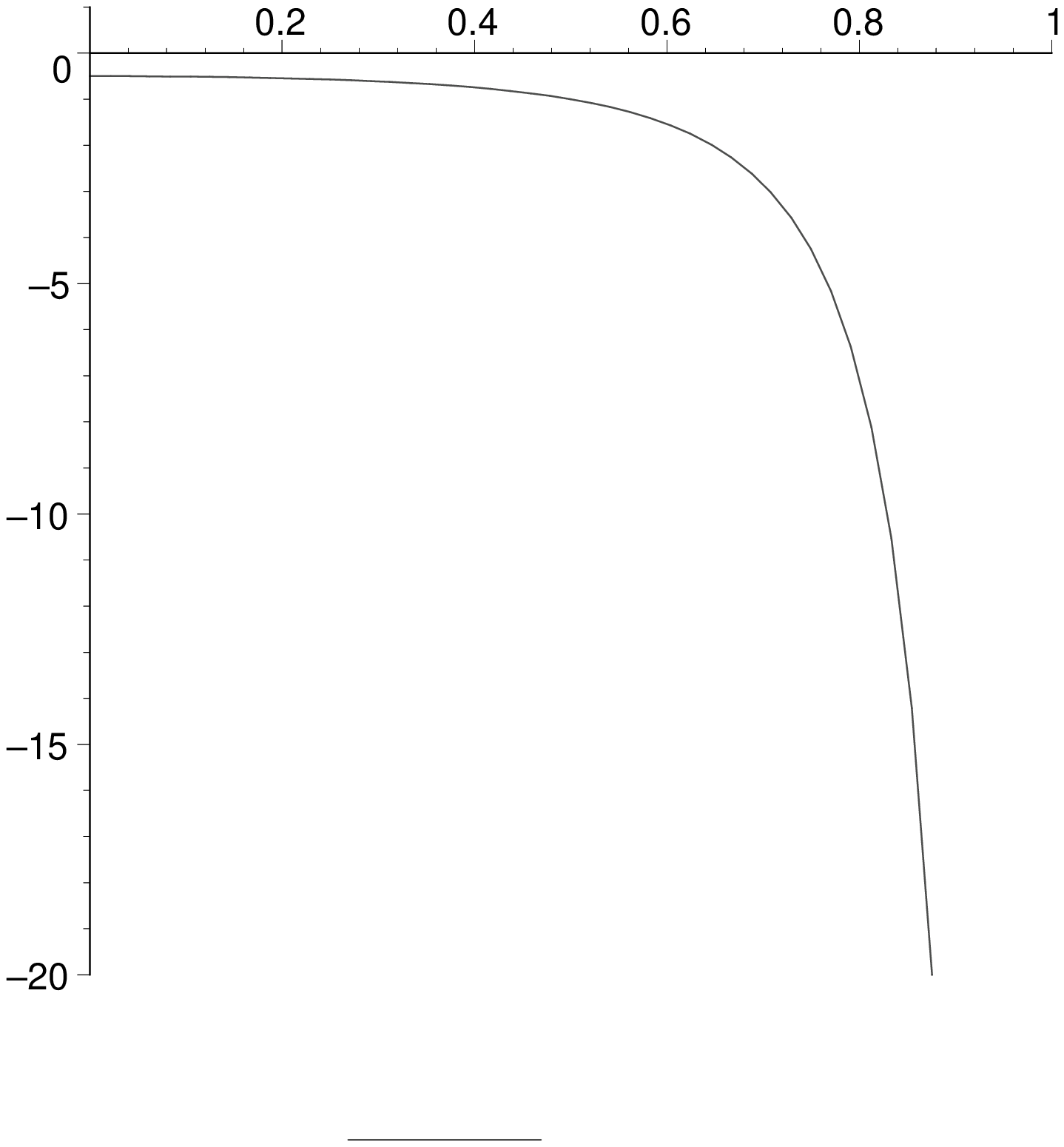}
\end{center}
\caption{The moduli potential as a function of
$\rho=a(t_2)/a(t_1)$ for ${\mathcal I}m T_{1,2}=0$,
 $ \vert w_2\vert \le \vert w_1 \vert $ and $\vert w_1\vert >
\vert Q\vert$.  The radion rolls down leading to colliding branes
with  an infinite negative energy. } \label{figure
potentiels 3}
\end{figure}

\begin{figure}[!h]
\begin{center}
\includegraphics[height=8cm]{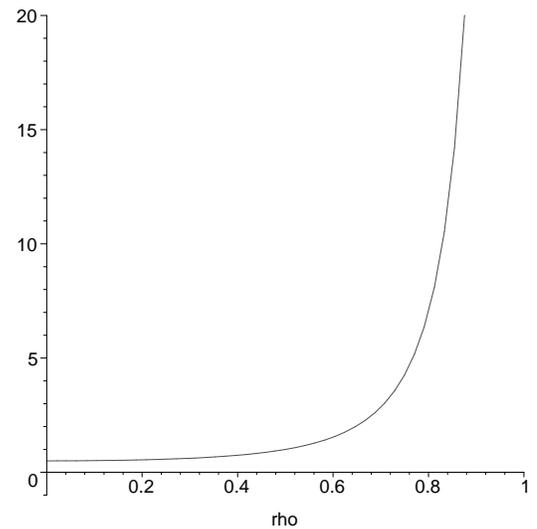}
\end{center}
\caption{The moduli potential as a function of
$\rho=a(t_2)/a(t_1)$ for ${\mathcal I}m T_{1,2}=0$,  $ \vert
w_2\vert \ge  \vert w_1 \vert$ and $\vert w_1 \vert  < \vert
Q\vert$. The radion is attracted towards a stable minimum with an
infinite interbrane distance, whose positive energy can be
fine-tuned to match the observed vacuum energy.} \label{figure
potentiels 4}
\end{figure}

\begin{figure}[!b]
\begin{center}
\includegraphics[height=8cm]{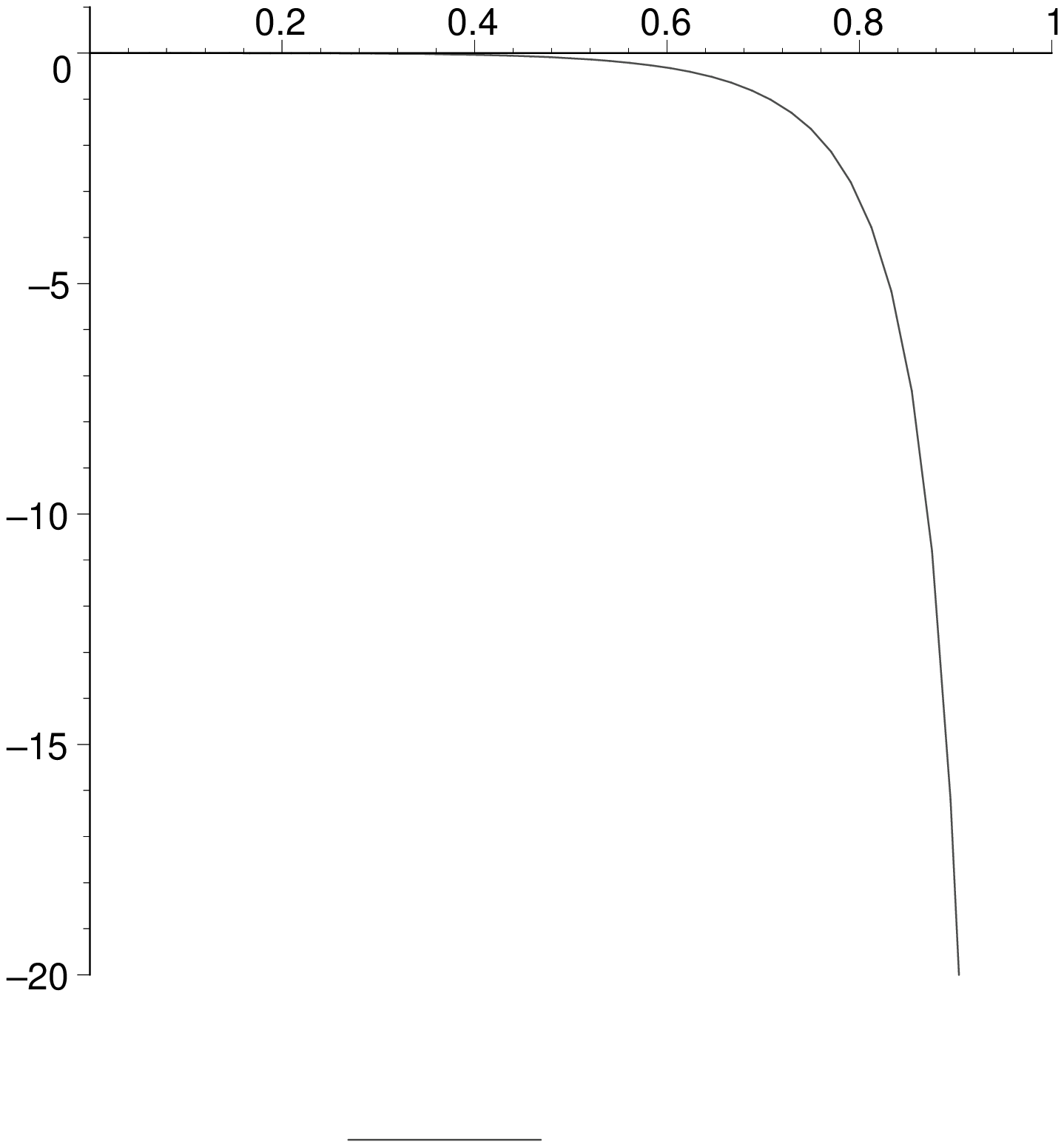}
\end{center}
\caption{The moduli potential as function of $\rho=a(t_2)/a(t_1)$
for ${\mathcal I}m T_{1,2}=0$ and  $ \vert w_2\vert <
\vert w_1 \vert = \vert Q\vert$.  The radion rolls down towards
colliding branes with an  infinite negative energy.}
\label{figure potentiels 5}
\end{figure}

\begin{figure}[!b]
\begin{center}
\includegraphics[height=8cm]{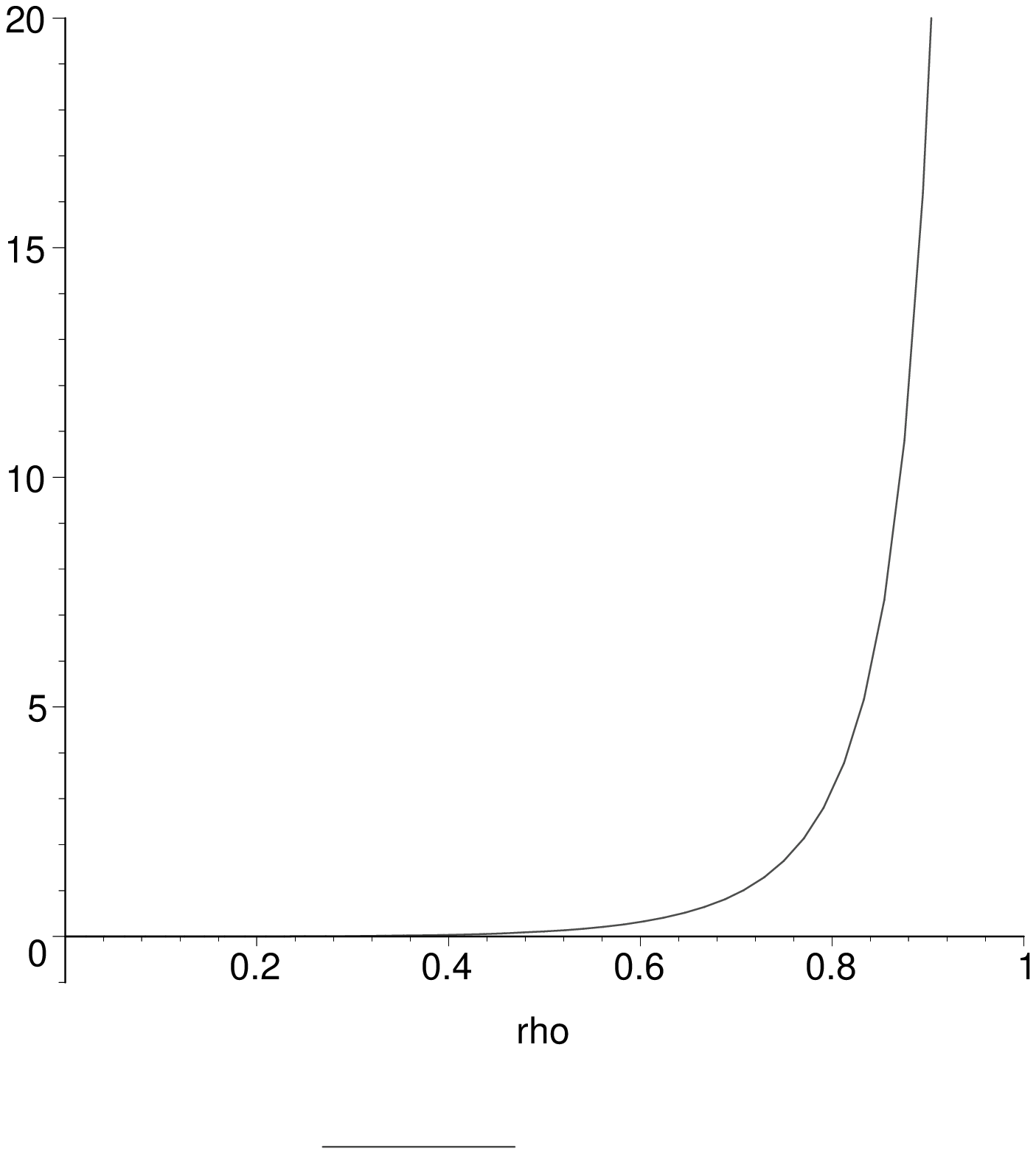}
\end{center}
\caption{The moduli potential as function of $\rho=a(t_2)/a(t_1)$
for ${\mathcal I}m T_{1,2}=0$
and  $ \vert w_2\vert > \vert w_1 \vert = \vert Q\vert$. There is a
minimum with zero energy  for an infinite interbrane distance.}
\label{figure potentiels 6}
\end{figure}

\subsection{Soft terms \label{soft terms section}}

Let us now discuss the soft  terms for general $w_i$.  We
restrict ourselves to the situation where matter is confined on
the \emph{positive tension brane}. This will allow us to evade
some of the gravitational constraints due to the presence of very
light moduli. We write the soft Lagrangian for the
\emph{canonically normalised} complex matter scalars $\tilde{s}_i$
as

\begin{equation}
{\cal L}_{soft}= -m^2_{i\bar j} \tilde{s}_i \overline{\tilde{s}_j} - (\frac{1}{2}B_{ij}\tilde{s}_i \tilde{s}_j + hc)
- (\frac{1}{6} A_{ijk}\tilde{s}_i \tilde{s}_j \tilde{s}_k + hc)
\end{equation}

Note that non vanishing soft terms do not necessarily imply
supersymmetry breaking (defined by non zero F--terms) when the
vacuum energy cannot be neglected, as can be seen from the general
expression of the soft masses (\ref{m2}) and B--terms (\ref{b}) ; we
will however keep the denomination of soft breaking terms for
simplicity.

Using the usual formulae for the soft breaking terms \cite{soft},
the un-normalised A--terms read
\begin{widetext}
\begin{eqnarray}
(A_{ijk})^{un.}&=&\frac{\bar W}{|W|}e^{K/2}F^{T_M}\Big[ \partial_{T_M} \lambda_{ijk}(T_p)
+K_{T_M} \lambda_{ijk}(T_p) -\Big((\partial_{T_M}K_{i\bar m}) K^{\bar mn}
\lambda_{njk}(T_p)
\nonumber \\
&&+ (i \leftrightarrow j) + (i \leftrightarrow k)\Big)\Big]
\nonumber \\
\Rightarrow A_{ijk} &=&12 \kappa_4^2 \lambda_{ijk} \alpha^2 (k {\mathcal I}m T_1)\Big( 2 \bar
w_2 a(\bar T_2)^{3-4\alpha^2} a(T_1)^{3-4\alpha^2} a(t_1)^{-6}k
{\mathcal I}m T_2
\nonumber \\
&&-\bar w_1 a(\bar T_1)^{6-4\alpha^2}a(T_1)^{-4\alpha^2} a(t_1)^{-6}
(\frac{\Delta a(t_1)^{-2}}{1+2\alpha^2}+2 i k {\mathcal I}m T_1) \Big) .
\end{eqnarray}
\end{widetext}
In this formula the Yukawa couplings $\lambda_{ijk}(T_p)=a^3(T_p)
\lambda_{ijk}$ are the moduli-dependent ones including the cubed
analytic warp factors, while $\lambda_{ijk}$ has no modulus
dependence. The notation will be the same for the next terms.
Notice that
\begin{equation}
A_{ijk}\equiv 0
\end{equation}
when the imaginary parts ${\cal I}m T_i$ vanish.
 This result is akin to
the vanishing of the $A$ terms in no--scale supergravity
\cite{noscale}, defined by the K\"ahler potential $K=-3 ln
\Big(\frac{\kappa_4^2}{\kappa_5^2}(T+\bar
T)-\frac{1}{3}\kappa_4^2\Sigma \bar \Sigma \Big)$. This can be
understood as in the limit of vanishing bulk curvature $k
\rightarrow 0$, the K\"ahler potential (\ref{kha}) takes the
no-scale form.

 The un-normalised soft masses for the scalars
\begin{widetext}
\begin{equation}
(m^2_{i\bar j})^{un.}= (m_{3/2}^2+\kappa_4^2<V>) K_{i\bar j}
-\overline{F^{T_M}}\Big(\partial_{\bar T_M}\partial_{T_N} K_{i\bar
j} - (\partial_{T_N} K_{i \bar k})K^{\bar k l}(\partial_{\bar T_M}
K_{l \bar j}) \Big)F^{T_N} \label{m2}
\end{equation}
\end{widetext}
are diagonal and read
\begin{widetext}
\begin{eqnarray}
m^2_{i \bar j} &=& \frac{2}{3}\kappa_4^2 <V> \delta_{i\bar j}+ 2
\alpha^2\kappa_4^4 \delta_{i\bar j}\Big|\frac{1}{1+2\alpha^2}\Delta^{-1/2}a(\bar
T_1 )^{3-4\alpha^2}a(t_1)^{-2}\bar w_1
\nonumber \\
&&+2\Delta^{-3/2}(k i{\mathcal I}m T_1)a(\bar
T_1)^{3-4\alpha^2}\bar w_1+2\Delta^{-3/2}(k i{\mathcal I}m
T_2)a(\bar T_2)^{3-4\alpha^2}\bar w_2\Big|^2
\nonumber \\
\end{eqnarray}
\end{widetext}
reducing to
\begin{widetext}
\begin{equation}
 m^2_{i\bar j}= \delta_{i \bar j}
\frac{2\kappa_4^4}{1+2\alpha^2}a(t_1)^{-12\alpha^2}\frac{|w_2|^2
\rho^{4-4\alpha^2}-|w_1|^2(1-\frac{\alpha^2}{1+2\alpha^2}(1-\rho^{2+4\alpha^2}))}{(1-\rho^{2+4\alpha^2})^2}
\end{equation}
\end{widetext}
for the canonically normalised matter fields living on the first
brane, and  for $Im T_i=0$. The supergravity vacuum energy
contribution $<V>$ to the soft masses is given in equation
(\ref{pot with im non zero}), or (\ref{pot}) for vanishing $Im
T_i$.

The general un-normalised effective $\mu$ term on the first brane is given by
\begin{equation}
(\mu^{eff}_{ij})^{un.}=\frac{\bar W}{|W|}e^{K/2}\mu_{ij}(T_p) + m_{3/2}\lambda_{ij}(T_p) - \overline{F^{T_M}}
\partial_{\bar T_M}\lambda_{ij}(T_p)
\end{equation}
where $\lambda_{ij}(T_p)=a^2(t_p) \lambda_{ij}$ are the
(moduli-dependent) Giudice-Masiero couplings given in
(\ref{kha_with_GM}), not to be confused with the Yukawa couplings
$\lambda_{ijk}$. The normalised effective $\mu$ term is
\begin{equation}
\mu^{eff}_{ij}=
<\frac{\overline{W}}{|W|}>\frac{a(T_1)^3}{a(t_1)^2\Delta^{1/2}}\Big(
\mu_{ij} +\lambda_{ij}\frac{\kappa_4^2
w_1a(T_1)^{-4\alpha^2}}{1+2\alpha^2}\Big)
\end{equation}
reducing to
\begin{equation}
\mu_{ij}^{eff}=
<\frac{\overline{W}}{|W|}>\frac{a(t_1)}{\Delta^{1/2}}\Big(
\mu_{ij} +\lambda_{ij}\frac{\kappa_4^2
w_1a(t_1)^{-4\alpha^2}}{1+2\alpha^2}\Big)
\end{equation}
for vanishing imaginary parts. Note that it does not depend on the
hidden brane detuning $w_2$, except non-relevantly through its
phase pre--factor.

 The general
un-normalised $B$ terms are
\begin{widetext}
\begin{eqnarray}
(B_{ij})^{un.}&=&\frac{\bar W}{|W|}e^{K/2}\Big[F^{T_M}\Big(\partial_{T_M}
\mu_{ij}(T_p) + K_{T_M}\mu_{ij}(T_p)-( \mu_{ik}(T_p)K^{k\bar l}\partial_{T_M}K_{j\bar l}
+ i \leftrightarrow j)\Big)
\nonumber \\
&& -m_{3/2}\mu_{ij}(T_p)\Big]+ (2m_{3/2}^2+\kappa_4^2 <V>)\lambda_{ij}(T_p) - m_{3/2}\overline{F^{T_M}}
\partial_{\bar T_M} \lambda_{ij}(T_p)
\nonumber \\
&& + m_{3/2}F^{T_M}\Big[\partial_{T_M}
\lambda_{ij}(T_p)-( \lambda_{ik}(T_p)K^{k\bar l}\partial_{T_M}K_{j\bar l}
+ i \leftrightarrow j)\Big]
\nonumber \\
&& - \overline{F^{T_M}}F^{T_N}\Big[\partial_{\bar T_M}
\partial_{T_N} \lambda_{ij}(T_p) -(K^{\bar k
l}\partial_{T_N}K_{i\bar k} \partial_{\bar T_M} \lambda_{lj}(T_p)
+ i \leftrightarrow j)\Big] \label{b}
\end{eqnarray}
\end{widetext}
In our case the normalised $B$ terms become
\begin{widetext}
\begin{eqnarray}
B_{ij}&=&
-\kappa_4^2(\mu^{eff}_{ij})_{\lambda=0}<\frac{W}{|W|}>\Big[\bar
w_1 \frac{a(\bar T_1)^{3-4\alpha^2}}{(1+2\alpha^2)a(t_1)^2
\Delta^{1/2}}
+\bar w_1 \frac{a(\bar T_1)^{3-4\alpha^2}}{a(t_1)^{4\alpha^2}
\Delta^{3/2}}(4 k \alpha^2 i {\mathcal I}m
T_1)\Big(\frac{3}{2\alpha^2}(4k\alpha^2 i{\mathcal I}m T_1)
\nonumber \\
&&+\frac{3}{1+2\alpha^2}(a(t_1)^{4\alpha^2}+a(T_1)^{4\alpha^2}+\Delta
a(t_1)^{-2} \Big)
+\bar w_2 \frac{a(\bar T_2)^{3-4\alpha^2}}{a(t_1)^{4\alpha^2}
\Delta^{3/2}}\frac{3}{2\alpha^2}(4k\alpha^2 i{\mathcal I}m
T_1)(4k\alpha^2 i{\mathcal I}m T_2)\Big]
\nonumber \\
&&+2\lambda_{ij}
\Delta^{-3}|w_2|^2 |a(T_2)|^{6-8\alpha^2}\Big[12\alpha^2 k^2({\cal I}m T_2)^2
+\frac{1}{1+2\alpha^2}a(t_2)^{-2+4\alpha^2}\Delta\Big]
\nonumber \\
&&+\lambda_{ij}[w_1\neq 0 \rm\ terms] .
\end{eqnarray}
\end{widetext}
For ${\mathcal I}m(T_i)=0$ we obtain
\begin{widetext}
\begin{eqnarray}
B_{ij} &=& -\Delta^{-2} a(t_1)\mu_{ij} \Big[ \kappa_4^2\bar w_1
a(t_1)^3\frac{1-\rho^{2+4\alpha^2}}{1+2\alpha^2}\Big]+ \lambda_{ij}
\kappa_4^2<V>
\nonumber \\
&&+ 2\lambda_{ij} \kappa_4^4 \Delta^{-3}{\cal R}e\Big(w_1
<\overline W>\Big)a(t_1)^3\frac{1-\rho^{2+4\alpha^2}}{1+2\alpha^2}
\nonumber \\
&& - \frac{\lambda_{ij}}{1+2\alpha^2} \kappa_4^4 \Delta^{-2}
a(t_2)^{2+4\alpha^2} |w_1 a(t_1)^{1-4\alpha^2}+w_2
a(t_2)^{1-4\alpha^2}|^2 .
\end{eqnarray}
\end{widetext}
We will give a
simplified expression when discussing the phenomenology of these
models. The effective Yukawa couplings are given by
\begin{equation}
\lambda_{ijk}^{eff}=<\frac{\overline{W}}{|W|}>\frac{a(T_1)^3}{a(t_1)^3}\lambda_{ijk}
\end{equation}
reducing to
\begin{equation}
\lambda_{ijk}^{eff} =<\frac{\overline{W}}{|W|}>\lambda_{ijk} .
\end{equation}
when the imaginary parts vanish.

Finally the classical gaugino masses for gauge fields living on
the first brane are
\begin{equation}
m_a=-\frac{k}{2} \sum_{i=1}^2 \beta_a^i a(t_i)^{-4\alpha^2}
F^{T_i} \label{gaugino masses}
\end{equation}
where we have defined $\beta_a^i=\frac{\partial \ln f_a}{\partial
\ln a(T_i) }$ depending on the gauge coupling function $f_a$ for
each gauge group $G_a$. As discussed in subsection \ref{soft terms
section}, the case $w_1=0$ (positive potential) will be the one
relevant phenomenologically. Considering that classically, five
dimensional locality implies $\beta^2_a=0$ (as discussed in
subsection \ref{section coupling}), we find that
\begin{equation}
m_a=0
\end{equation}
as long as ${\cal I}m T_i=0$. There are different possibilities to
generate non zero gaugino masses. For example one may add a new
source of supersymmetry breaking due to a new field, like in the
next section ; later in section \ref{AMSB} we will also consider
the contribution of anomaly mediated supersymmetry breaking to the
gaugino masses. In the following section we will explore the
possibilities opened up by the presence of a dilaton in order to
generalise our construction.

\section{Supersymmetry Breaking and the Dilaton}

\subsection{The moduli potential}

So far we have concentrated on the case where only one vector
multiplet is present in the bulk. In this section we will focus on
the case where both a vector multiplet and the universal
hypermultiplet are present. The nature of the scalar potential in
that case changes drastically. It becomes very intricate to study
the stability of the possible extrema.

We will consider the moduli K\"ahler potential
\begin{equation}
K=-\ln(S+\bar S) - 3 \ln\Big(a(\frac{T_1+\bar
T_1}{2})^{2+4\alpha^2}-a(\frac{T_2+\bar
T_2}{2})^{2+4\alpha^2}\Big)
\end{equation}
and an arbitrary superpotential first
\begin{equation}
W=W(T_1,T_2,S)
\end{equation}
where we have suppressed the dependence on the matter fields. The
scalar potential reads now
\begin{widetext}
\begin{eqnarray}
V&=&\frac{|(S+\bar S)W_S - W|^2}{(S+\bar
S)\Delta^3}+\frac{|W_{T_2}|^2 a(t_2)^{-2+4\alpha^2}-|W_{T_1}|^2
a(t_1)^{-2+4\alpha^2}}{3(1+2\alpha^2)k^2(S+\bar S)\Delta^2}
\nonumber \\
&&+
\frac{|W_{T_1}a(t_1)^{4\alpha^2}+W_{T_2}a(t_2)^{4\alpha^2}+3kW|^2}{6\alpha^2k^2(S+\bar
S)\Delta^3} .
\end{eqnarray}
\end{widetext}
 Notice the
similarity with case when no dilaton is present. Choosing the
superpotential to be of the form
\begin{equation}
W=w_1(S)a(T_1)^3+w_2(S)a(T_2)^3
\end{equation}
where the two functions $w_i(S)$ replace the constant
superpotentials $w_i$ of the previous section, we find that
\begin{widetext}
\begin{eqnarray}
V&=&\frac{|(S+\bar S)W_S - W|^2}{(S+\bar S)\Delta^3}
\nonumber \\
&&+\frac{3\kappa_4^2}{(1+2\alpha^2)(S+\bar S)\Delta^2}\Big[ \vert
a(T_2)^{3-4\alpha^2} w_2(S)\vert^2 a(t_2)^{-2+4\alpha^2}-\vert
a(T_1)^{3-4\alpha^2} w_1(S)\vert^2 a(t_1)^{-2+4\alpha^2}\Big]
\nonumber \\
&&+\frac{24\alpha^2 k^2 \kappa_4^2}{(S+\bar S)\Delta^3}\vert
a(T_1)^{3-4\alpha^2} w_1(S) {\cal I}m T_1 +a(T_2)^{3-4\alpha^2}
w_2(S) {\cal I}m T_2\vert^2.
\end{eqnarray}
\end{widetext}
The structure of the potential when $w_i(S)$ are not constant is
very difficult to analyse. In the following we will study the case
where only one of the terms $w_2(S)$ is present.

\subsection{ Race--track models}

Let us assume that strong coupling effects lead to a potential for
the dilaton on the second brane. Including matter fields the
superpotential reads now
\begin{equation}
W=a(T_1)^3(\frac{1}{2}\mu \Sigma^2 + \frac{1}{6}y\Sigma^3)
+a(T_2)^3 w_2 h(S) \label{racetrack W}
\end{equation}
where $h(S)$ results from strong gauge coupling effects. In that
case the scalar potential for the moduli reads
\begin{equation}
V = \kappa_4^{-4} e^{\hat G}\frac{|(S+\bar S)h'(S)-h(S)|^2}{S+\bar
S} +\frac{|h(S)|^2}{S+\bar S}\hat V .
\end{equation}
where hatted quantities refer to the case with no dilaton. This
potential admits extrema in $S$ satisfying
\begin{widetext}
\begin{equation}
\Big((S+\bar S)\overline{h'(S)}-\overline{h(S)}\Big) h''(S)(S+\bar
S)^2 =-\Big((S+\bar S)h'(S)-h(S)\Big)\overline{h(S)}(1+\kappa_4^4
e^{-\hat G} \hat V) .
\end{equation}
\end{widetext}
There are two types of extrema. First of all when
\begin{equation}
(S+\bar S)h'(S)=h(S)
\end{equation}
we find that
\begin{equation}
F^S=0
\end{equation}
Notice that $S$ is determined independently of the other moduli.
In that case the potential reduces to
\begin{equation}
V=\frac{|h(S)|^2}{(S+\bar S)}\hat V
\end{equation}
at the extremum. The other extrema satisfy the necessary condition
\begin{equation}
|\frac{h''(S)}{h(S)}|(S+\bar S)^2 = |1+\kappa_4^4 e^{-\hat G}\hat
V| .
\end{equation}
Let us concentrate on the gaugino condensation case (see
\cite{gaugino condensate} for a recent review) where
\begin{equation}
h(S)=\exp(-\frac{3}{2b_0}S) \label{racetrack W 2}
\end{equation}
The potential has then either no extremum or a maximum, i.e. the
dilaton has a run--away potential, see figures 7 and 8.


\begin{figure}
\begin{center}
\includegraphics[height=8cm]{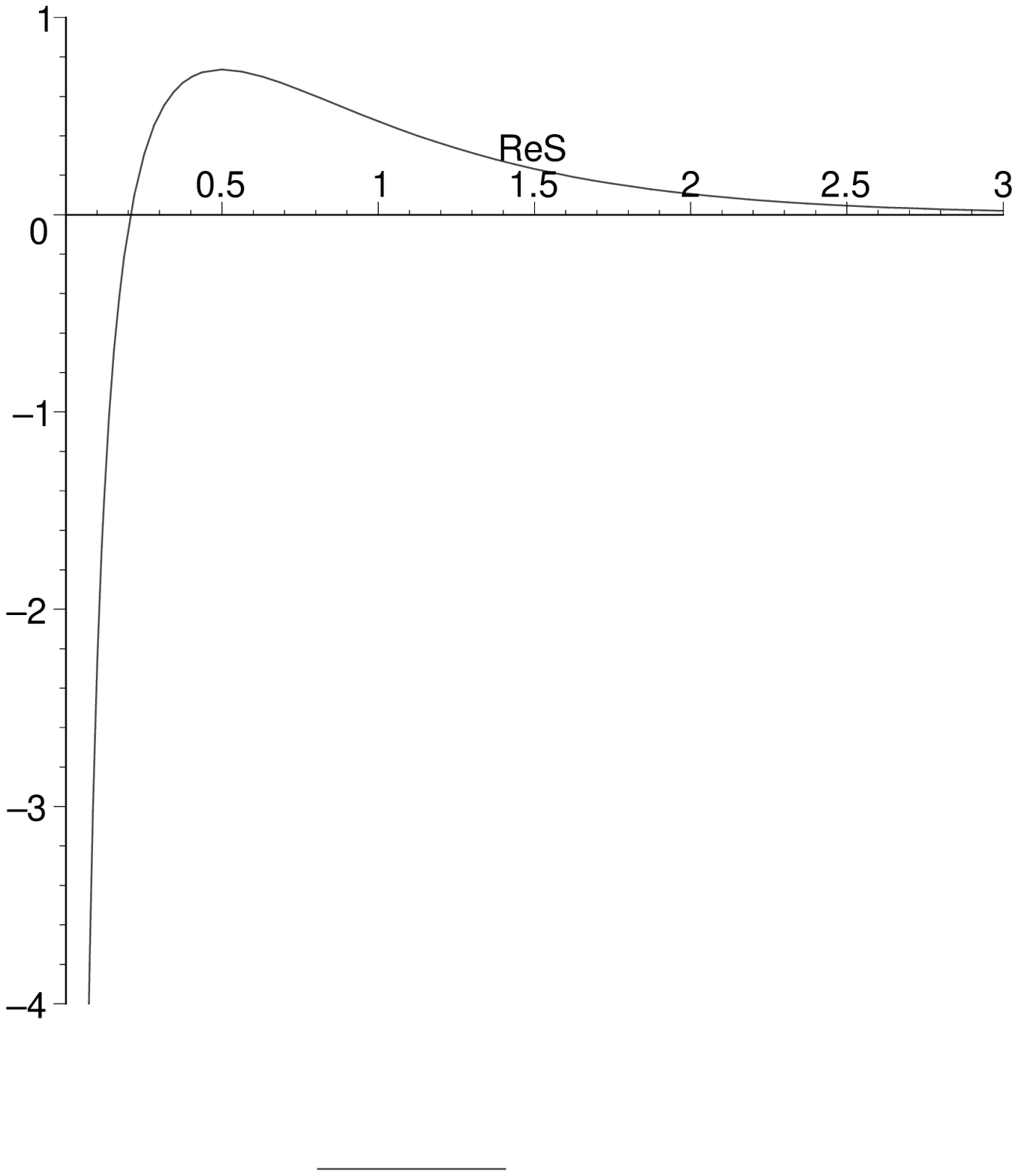}
\end{center}
\caption{The moduli potential as a function of the dilaton $Re(S)$
with detunings $w_1=0$ and $w_2=exp(-\frac{3}{2b_0}S)$ (gaugino
condensation on the hidden brane), depending on the value of $\hat
V(T_i)$ and $\hat G(T_i)\equiv \hat K(T_i) + ln|\kappa_4^{3}\hat
W(T_i)|^2$. Notice that $\kappa_4^4 \hat V + e^{\hat G}$ is
negative.}
\end{figure}

\begin{figure}
\begin{center}
\includegraphics[height=8cm]{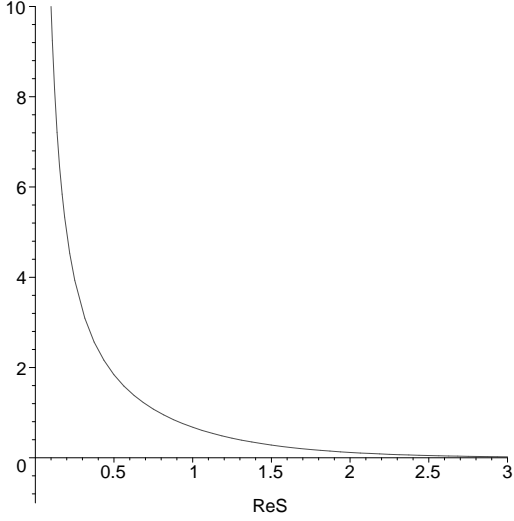}
\end{center}
\caption{The moduli potential as a function of the dilaton $Re(S)$
with detunings $w_1=0$ and $w_2=exp(-\frac{3}{2b_0}S)$ (gaugino
condensation on the hidden brane), depending on the value of $\hat
V(T_i)$ and $\hat G(T_i)\equiv \hat K(T_i) + ln|\kappa_4^{3}\hat
W(T_i)|^2$. Notice that  $\kappa_4^4 \hat V + e^{\hat G}$ is
 positive.}
\end{figure}

Another relevant case is the race--track superpotential on the
second brane where
\begin{equation}
h(S)=\lambda_1 \exp(-b_1 S) + \lambda_2 \exp(-b_2 S)
\end{equation}
with $b_1,b_2 \geq 0$. This factor originates from the gaugino
condensation in two different gauge groups, and is especially
motivated in our case as being known to allow for stabilization
\cite{racetrack}. The extrema with $F^S=0$ satisfy
\begin{eqnarray}
\Big|\frac{\lambda_2}{\lambda_1}\Big|\exp\Big((b_1-b_2)Re(S)\Big)=\frac{1+2b_1
Re(S)}{1+2b_2 Re(S)}
\nonumber \\
(b_1-b_2)Im(S)= \pi +arg(\frac{\lambda_1}{\lambda_2})\ (mod\ 2\pi)
\end{eqnarray}
with a unique solution in Re(S), at least when
$|\frac{\lambda_2}{\lambda_1}|=1$. The stability of this
configuration depends on the moduli $T_i$ and deserves further
study.

The soft terms can be related to the soft terms when no dilaton is
present. One can of course evaluate them at the various extrema.
\begin{widetext}
\begin{eqnarray}
y_{eff} &=& (S+\bar S)^{-1/2}\hat y_{eff}
\nonumber \\
\mu_{eff} &=& (S+\bar S)^{-1/2}\hat \mu_{eff}^{no\ GM} +
|h(S)|(S+\bar S)^{-1/2}\hat \mu_{eff}^{GM\ only}
\nonumber \\
m_{3/2} &=&  |h(S)|(S+\bar S)^{-1/2} \hat m_{3/2}
\nonumber \\
m^2 &=&|h(S)|^2(S+\bar S)^{-1} \hat m^2 +\kappa_4^{-2}
\frac{|F^S|^2}{(S+\bar S)^2}
\nonumber \\
A &=& |h(S)|(S+\bar S)^{-1} \hat A
\nonumber \\
B &=& |h(S)|(S+\bar S)^{-1} \hat B^{no\ GM} +
\lambda_{GM}\kappa_4^{-2} \frac{|F^S|^2}{(S+\bar
S)^2}+|h(S)|^2(S+\bar S)^{-1} \hat B^{GM\ only}
\nonumber \\
m_a &=& |h(S)|(S+\bar S)^{-1/2}\hat m_a \label{dilaton rescaling}
\end{eqnarray}
\end{widetext}

For most terms, this results in a simple rescaling with
S-dependent factors. Note that accordingly the Giudice-Masiero
parts of the effective $\mu$-term and the B-term are not rescaled
like their ordinary parts. The soft masses and B-terms also have
new additive contributions in $|F^S|^2$ from their vacuum energy
term.

To conclude the racetrack case, we find that as long as the
imaginary parts of the $T_i$ moduli vanish and $S$ sits at the
extremum where $F^S=0$, the gauginos are still massless $m_a=0$.
In section \ref{AMSB} we will see  however that the gauginos can
pick up a mass via anomaly mediation.

\section{Anomaly Mediated Supersymmetry Breaking \label{AMSB}}

We have seen that the tree level action does not mix the moduli
dependence of the gauge coupling functions coming from each brane.
This is due to the locality of the coupling to the bulk scalar
field in five dimensions. Now there is a quantum conformal anomaly
at one loop which leads to a coupling of both moduli to the gauge
sectors on each brane. We will follow closely the superfield
method of \cite{anomaly} in order to derive its consequences on
the soft breaking terms, especially the gaugino masses and the $A$
terms.

Let us consider the supergravity action written with the chiral
compensator formalism. To simplify the discussion we only consider
a single matter $\Sigma$ field on the first brane
\begin{widetext}
\begin{eqnarray}
&\int d^4x d^2 \theta d^2 \bar\theta\ \Big[-3 f(t_1,t_2)+a^2(t_1)
\Big(|\Sigma|^2+\frac{1}{2}(\lambda\Sigma^2+\bar \lambda\bar
\Sigma^2)\Big)\Big]|\Phi|^2 &
\nonumber \\
&+\int d^4x d^2\theta\ \Phi^3 a^3(T_1)[\frac{1}{2}\mu \Sigma^2 +
\frac{1}{6}y\Sigma^3] + h.c.& .
\end{eqnarray}
\end{widetext}
This is the action in the brane frame as can be seen from the
non-canonical term $-3 f(t_1,t_2)|\Phi|^2$.  The coupling to the
moduli has been determined in section 2. We have included a
Giudice-Masiero term involving the coupling $\lambda$. The
$\theta=0$ component of the conformal compensator has not yet been
gauge-fixed.

In the following we will factorise real superfields $R$
\begin{equation}
R(\theta,\bar \theta)= R_0 +R_1 \theta^2 +\bar R_1 \bar \theta^2
+R_2 \theta^2 \bar\theta^2
\end{equation}
as
\begin{equation}
R(\theta,\bar\theta)= R_0(1+\frac{R_1}{R_0}\theta^2) (1+\frac{\bar
R_1}{\bar R_0}\bar\theta^2)(1 +D_R\theta^2\bar\theta^2)
\end{equation}
where $D_R=\frac{R_2}{R_0}-\frac{|R_1|^2}{R_0^2}$. We will denote
by
\begin{equation}
{\cal {R}}=R_0^{1/2}(1+\frac{R_1}{R_0}\theta^2)
\end{equation}
the chiral part of the factorisation of $R(\theta,\bar\theta)$.

The change to the superspace Einstein frame is realized by the
chiral superfield redefinition
\begin{equation}
\tilde{\Phi}={\mathcal F} \Phi
\end{equation}
where ${\cal F}=f^{1/2}|_{\theta=0}(1+\theta^2
\frac{1}{2}\frac{f_{t_i}}{f}F^{T_i})$.
 The action becomes
\begin{widetext}
\begin{eqnarray}
&\int d^4x d^2 \theta d^2 \bar \theta\
\Big[-3 (1+D_f\theta^2\bar\theta^2)|\tilde\Phi|^2+\frac{|\tilde\Phi|^2}{|{\mathcal
F}|^2}a^2(t_1) \Big(|\Sigma|^2+\frac{1}{2}(\lambda\Sigma^2+\bar
\lambda\bar \Sigma^2)\Big)\Big]&
\nonumber \\
&+\int d^4x d^2\theta\ \frac{\tilde \Phi^3}{{\mathcal F}^3}
a^3(T_1)[\frac{1}{2}\mu \Sigma^2 + \frac{1}{6}y\Sigma^3] + h.c.&
\end{eqnarray}
\end{widetext}
The $D_f$ term contributes to the classical action only. One can
now gauge-fix $\tilde{\Phi}|_{\theta=0}=1$ corresponding to the
Einstein frame. Let us also factorise the real superfield
$a^2(t_1)$.
\begin{widetext}
\begin{eqnarray}
&\int d^4x d^2 \theta d^2 \bar \theta\  \Big[-3|\tilde
\Phi|^2(1+D_f\frac{1}{4}\theta^2\bar\theta^2)] +\frac{|\tilde
\Phi|^2 |{\cal A}|^2}{|{\cal F}|^2}(1+ D_{a^2}\theta^2\bar
\theta^2) \Big(|\Sigma|^2+\frac{1}{2}(\lambda\Sigma^2+\bar
\lambda\bar \Sigma^2)\Big)\Big]&
\nonumber \\
&+\int d^4x d^2\theta\ \frac{\tilde{\Phi}^3 a^3(T_1)}{{\mathcal
F}^3}[\frac{1}{2}\mu \Sigma^2 + \frac{1}{6}y\Sigma^3] + h.c.&
 .
\end{eqnarray}
\end{widetext}
where ${\cal A}=a(t_1)|_{\theta=0}(1+\frac{\partial \ln
a(t_1)}{\partial t_i} F^{T_i})$. Now we can redefine the matter
fields
\begin{equation}
\tilde \Sigma= \frac{\tilde \Phi{\cal A}}{\cal F} \Sigma\equiv
{\cal G}\Sigma
\end{equation}
Explicitly this reads
\begin{equation}
\tilde \Sigma =\frac{a(t_1)}{f^{1/2}}\Big|_{\theta=0} (1+F_{\cal
G}\theta^2)\Sigma
\end{equation}
where
\begin{equation}
F_{\cal G}= (F_{\tilde\Phi}+\frac{\partial \ln\
a(t_1)f^{-1/2}}{\partial t_i}F^{T_i}).
\end{equation}
Notice that according to this definition
of $F_{\cal G}$, the $F$-term of $\cal{G}$ is actually $\frac{a(t_1)}{f^{1/2}}\Big|_{\theta=0}F_{\cal G}$.
 The matter
action becomes
\begin{widetext}
\begin{eqnarray}
&\int d^4x d^2 \theta d^2 \bar\theta
\Big(1+\theta^2\bar
\theta^2|F^{T_1}|^2\frac{1}{a(t_1)}\frac{\partial^2
a(t_1)}{(\partial
t_1)^2}\Big)\Big(|\tilde\Sigma|^2+\frac{1}{2}\lambda
\frac{\bar{\mathcal G}}{\mathcal G}\tilde\Sigma^2+h.c.\Big)&
\nonumber \\
&+\int d^4x d^2\theta\ \frac{\tilde{\Phi}^3 a^3(T_1)}{{\mathcal
F}^3{\mathcal G}^3} [\frac{1}{2}\mu{\mathcal G} \tilde \Sigma^2 +
\frac{1}{6}y\tilde\Sigma^3] + h.c.&
\label{classical superspace soft br action}
\end{eqnarray}
\end{widetext}
This is the classical action in terms of the normalised matter
fields in the Einstein frame. The superpotential action can be rewritten as
\begin{equation}
\int d^4x d^2\theta\ \frac{a^3(T_1)}{{\mathcal A}^3} [\frac{1}{2}\mu{\mathcal G} \tilde \Sigma^2 +
\frac{1}{6}y\tilde\Sigma^3] + h.c.
\label{classical superspace superpot rewritten}
\end{equation}
The field redefinitions that we have
performed are all anomalous. Let us now write the action with
renormalized couplings, evaluated at a given energy scale $E$. The
superpotential couplings are not renormalised. The effective
ultra-violet  field-dependent cut-off  is $\Lambda_{UV}{\mathcal
G}$ and is superfield dependent. This superfield ${\mathcal G}$
breaking supersymmetry when $F_{\mathcal G}\ne 0$, the rescaling
generates anomalous soft terms. We thus have the kinetic part of
the action
\begin{widetext}
\begin{equation}
 S_{qu}\supset \int
d^4x d^2 \theta d^2 \bar \theta
\Big(Z(\frac{E}{\Lambda_{UV}|{\mathcal
G}|})|\tilde\Sigma|^2+\frac{1}{2}\lambda(\frac{E}{\Lambda_{UV}|{\mathcal
G}|})\tilde\Sigma^2+h.c.\Big)
\end{equation}
\end{widetext}
where we have discarded the $\theta^2\bar \theta^2 |F^{T_1}|^2\frac{1}{a(t_1)}\frac{\partial^2
a(t_1)}{(\partial
t_1)^2}$ and $\frac{\bar{\mathcal G}}{{\mathcal G}}|_{\theta\ne 0}$ terms of (\ref{classical superspace soft br action})
as contributing only to the tree level soft breaking action, already computed in the previous sections, and not to the anomalous terms. Similarly, in the classical superpotential action (\ref{classical superspace superpot rewritten}), the $\frac{a^3(T_1)}{{\mathcal A}^3}|_{\theta \ne 0}$ and ${\mathcal G}|_{\theta \ne 0}$ terms will be eliminated from now on.

Now the wave function
normalization Z contains $\theta^2$ and $\bar \theta^2$ terms
\begin{widetext}
\begin{equation}
Z(\frac{E}{\Lambda_{UV}|{\mathcal G}|})=Z(\frac{E
f^{1/2}(t_1,t_2)}{\Lambda_{UV}a(t_1)})\Big|_{\theta=0}\Big|1-\frac{\gamma}{2}\theta^2F_{\mathcal
G}\Big|^2 \Big(1+\frac{1}{8}\theta^2\bar \theta^2(\frac{\partial
\gamma}{\partial g}\beta_g+\frac{\partial \gamma}{\partial \ln
\lambda}\xi)|F_{\mathcal G}|^2\Big)
\end{equation}
We can redefine the matter fields using
\begin{equation}
\hat \Sigma=
Z^{1/2}|_{\theta=0}\Big(1-\frac{\gamma}{2}\theta^2F_{\mathcal
G}\Big)\tilde\Sigma .
\end{equation}
\end{widetext}
where we have defined $\gamma\equiv \frac{\partial ln Z}{\partial
ln E}$ the anomalous dimension of $\tilde \Sigma$, similarly $\xi
\equiv \frac{\partial ln \lambda}{\partial ln E}$, and
$\beta_g\equiv \frac{\partial g}{\partial ln E}$ where $g$ is a
gauge coupling. The resulting action involves several blocks each
leading to soft breaking terms. The soft masses come from
\begin{equation}
\int d^4x d^2 \theta d^2 \bar \theta\ (1-\theta^2\bar \theta^2
m^2_{an.})|\hat\Sigma|^2
\end{equation}
and read
\begin{equation}
m^2_{an.,\bar i j}=-\frac{1}{8}\delta_{\bar i j}(\frac{\partial
\gamma_{i}}{\partial g}\beta_g+\frac{\partial \gamma_{i}}{\partial
\ln \lambda_{kl}}\xi_{kl})|F_{\mathcal G}|^2
\end{equation}
We have restored the index structure in the final formulae to take
into account the case with  several chiral multiplets, with the
simplification assumption of diagonal wave function
renormalisation $Z_{\bar i j}=Z_i \delta_{\bar i j}$. Notice that
this is the generalisation of the result of Randall and Sundrum
where the contribution from the moduli fields has been taken into
account. The $B$ and the $\mu$ terms are also generated at one
loop
\begin{eqnarray}
&\int d^4x d^2 \theta d^2\bar \theta \frac{1}{2}\Big((\lambda_R+
\bar\theta^2\mu_{an.}-\theta^2\bar\theta^2
B_{an.}^{G.M.})\hat\Sigma^2+h.c.\Big)&
\nonumber \\
&+\int d^4x d^2\theta \frac{1}{2}(\mu_R +\theta^2 B^\mu_{an.})\hat
\Sigma^2 + h.c.&
\end{eqnarray}
where $ \lambda_{R, ij}= \frac{\lambda_{ij}}{\sqrt{Z_iZ_j}}$ and
are given by
\begin{equation}
B_{an, ij}= B^{\mu}_{an, ij}+B_{an, ij}^{G.M.}
\end{equation}
with
\begin{equation}
B_{an., ij}^\mu=\frac{\gamma_i+\gamma_j}{2}  F_{\mathcal G}\mu_{R,
ij}
\end{equation}
where $\mu_{R, ij}=
\frac{a^3(T_1)}{a^2(t_1)f^{1/2}\sqrt{Z_iZ_j}}\mu_{ij}$ and
\begin{equation}
B_{an., ij}^{G.M.}= \frac{1}{4}\Big(\frac{1}{2}\frac{\partial
\xi_{ij}}{\partial g}\beta_g+\frac{1}{2}\frac{\partial
\xi_{ij}}{\partial \ln
\lambda_{kl}}\xi_{kl}-(\gamma_i+\gamma_j)\xi_{ij}\Big)|F_{\mathcal
G}|^2\lambda_{R, ij}
\end{equation}
The anomalous effective $\mu$ term is given by
\begin{equation}
 \mu_{an., ij}=- \frac{\xi_{ij}}{2} F_{\mathcal G} \lambda_{R, ij}.
\end{equation}
Notice that when the tree level $\mu_{eff}=0$, one can generate a
loop contribution to the $\mu$ term via the Giudice--Masiero term.
Finally the $A$ terms are also generated at one loop from
\begin{equation}
\int d^4x d^2\theta \frac{1}{6}(y_R+\theta^2 A_{an.})\hat\Sigma^3
+ h.c.
\end{equation}
where $y_{R,ijk}=
\frac{a^3(T_1)}{a^3(t_1)\sqrt{Z_iZ_jZ_k}}|_{\theta=0}y_{ijk}$
leading to
\begin{equation}
A_{an., ijk}=\frac{\gamma_i+\gamma_j +\gamma_k}{2}F_{\mathcal
G}y_{R,ijk}
\end{equation}
The order parameter for anomaly mediation is $F_{\mathcal G}$. The
contribution in $F^{T_i}$ arising in $F_{\mathcal G}$ springs from
the conformal anomaly in going from the brane frame to the
Einstein frame. We have already computed $F^{T_i}$, we find now
\begin{widetext}
\begin{equation}
F_{\tilde \Phi}=\frac{2\kappa_4^2}{3\Delta^{3/2}}\Big(ik{\cal I}m
T_1 a(\bar T_1)^{3-4\alpha^2} \bar w_1 +ik{\cal I}m T_2 a(\bar
T_2)^{3-4\alpha^2} \bar w_2 \Big)
+\frac{1}{2}\frac{f_{t_i}}{f}F^{T_i}
\end{equation}
\end{widetext}
This can be obtained by noticing that the superspace volume
element $E^{-1}\supset \bar \Phi \Phi$.

Note that in principle the running couplings are now evaluated for
a cut-off
\begin{equation}
\Lambda_{UV}\Big|1-\frac{\gamma}{2}\theta^2 F_{\mathcal G}\Big|
\end{equation}
which is again $\theta$-dependent. This dependence should be
expanded, yielding extra contributions to the soft terms, but
these would be of higher order and were consequently neglected.

Let us conclude with the anomalous gaugino masses, deduced from
the running gauge coupling function
\begin{widetext}
\begin{equation}
f(E)=f^0 + \frac{b}{2\pi} \ln(\frac{E}{\Lambda_{UV}{\mathcal G}})
= f_i^0 +  \frac{b}{2\pi} \ln(\frac{E
f^{1/2}(t_1,t_2)}{\Lambda_{UV}a(t_1)}|_{\theta=0}) -
\frac{b}{2\pi}\theta^2 F_{\mathcal G} .
\end{equation}
\end{widetext}
Thus the anomalous contribution to the gaugino masses is
\begin{equation}
m^{an.}_{a}=\frac{b_a}{2\pi}F_{\mathcal G} .
\end{equation}
Notice that all the soft terms emanating from anomaly mediation
are of order $F_{\mathcal G}$.

\section{Soft Supersymmetry Breaking from the Hidden Brane}

We come back to the case where no hypermultiplet is in the bulk.
Having analysed the breaking of supersymmetry in the brane models
with a vector multiplet scalar field, we will extract ingredients
with phenomenological relevance. We will focus on the case where the
potential admits a global minimum for $\rho=0$, $w_1=0$, i.e.
supersymmetry breaking is only due to the negative tension brane.
The axions will be taken to be at  the extremum ${\mathcal I}m
T_i=0$, which in this case is stable for ${\mathcal I}m T_2$ and
flat for ${\mathcal I}m T_1$. Notice that $F^{T_2}\ne 0$ implying that supersymmetry is broken.

The supersymmetry breaking can be better understood using the
normalised moduli fields. We redefine the non-axionic fields in the following
way:
\begin{eqnarray}
a(t_1)^{1+2\alpha^2} &=& e^\sigma \cosh r
\nonumber \\
a(t_2)^{1+2\alpha^2} &=& e^\sigma \sinh r
\end{eqnarray}
Notice that $r=0$ when the second brane hits the singularity while
$r=\infty$ when the two branes collide. The sliding field $\sigma$
describes the position of the centre of mass of the two branes. In
the Einstein frame the kinetic terms are expressed as
\begin{equation}
S=\frac{1}{2\kappa_4^2}\int d^4 x \sqrt {-g}( {\cal R}^{(4)}
-\frac{12\alpha^2}{1+2\alpha^2}(\partial \sigma)^2
-\frac{6}{1+2\alpha^2} (\partial r)^2)
\end{equation}
This is a sigma--model Lagrangian with normalisation matrix
$\gamma_{rr}=\frac{6}{1+2\alpha^2}$ and
$\gamma_{\sigma\sigma}=\frac{12\alpha^2}{1+2\alpha^2}$. Notice
that $\sigma$ decouples in the RS model as the centre of mass of
the brane system is irrelevant.  Moreover in the RS model,
\begin{equation}
\tanh r=e^{-kt}
\end{equation}
where $t$ is the radion.

Before analysing the structure of the soft terms, let us
concentrate on the $\mu$ problem. Indeed the $\mu$ term must be of
the order of the weak scale. There are two sources for the $\mu$
term here, the supersymmetric $\mu$ term in the superpotential and
the Giudice-Masiero term in the K\"ahler potential. It turns out
that one cannot use the Giudice-Masiero mechanism to obtain a
small $\mu$ term indeed
\begin{equation}
\mu^{eff}\approx \mu e^{-\frac{2\alpha^2}{1+2\alpha^2}\sigma}
\end{equation}
is independent of $\lambda_{GM}$. However it is possible to use
the field $\sigma$ to generate the electroweak hierarchy between a
Planck scale fundamental $\mu$-term and the weak scale effective
$\mu$-term. This new mechanism using the sliding field $\sigma$ is
not possible in the pure Randall-Sundrum model with $\alpha=0$.

Another interesting scenario can be obtained when the classical
fundamental $\mu$ term vanishes. At one loop the $\mu$ term
depends on $F_{\mathcal G}$
\begin{equation}
F_{\mathcal G}  \approx 2 \kappa_4^2 \bar w_2
e^{-\frac{6\alpha^2\sigma}{1+2\alpha^2}}r^{\frac{3}{1+2\alpha^2}}
\end{equation}
We thus obtain that
\begin{equation}
\mu=-\frac{\xi}{2} \lambda_{GM}  F_{\mathcal G}
\end{equation}
with $\xi\equiv \frac{\partial ln \lambda_{GM}}{\partial ln E}$.
The $\mu$ term can be of the order of the electro--weak scale
provided the $F_{\mathcal G}$ breaking term is in TeV range.

The soft masses are given by
\begin{equation}
m^2 \approx \frac{1}{\lambda_{GM}}B
\end{equation}
where the tree level contribution dominates. Hence the $B$ terms
are naturally of order of the soft masses. Now the soft masses are
related to $F_{\mathcal G}$ breaking term via
\begin{equation}
m^2
 \approx
\frac{1}{2(1+2\alpha^2)} \frac{|F_{\mathcal G}|^2}{ r^{2}} .
\end{equation}
Notice that for small $r$ the soft masses can be larger than the
TeV range. Of course this can be modified provided $\xi \lambda$
is large enough. In that case $\mu\approx \xi \lambda F_{\mathcal
G}\approx F_{\mathcal G}/r\approx m$ may both be at the
electro--weak scale even for small $r$, at the cost of a unnatural
relation between the parameters $r_{now}$ and $\xi\lambda$.

Supersymmetry is broken and the gravitino mass becomes
\begin{equation}
m_{3/2} \approx \frac{F_{\mathcal G}}{2}
\end{equation}
implying that the gravitino mass can also be within the $TeV$
range.

Now the tree level gaugino masses vanish as ${\cal I}m T_i=0$. A
non-vanishing contribution results from anomaly mediation.
Therefore the anomalous contribution becomes
\begin{equation}
m^{an.}_a \approx \sqrt{\frac{1+2\alpha^2}{2}}\frac{b_a}{\pi} m
r
\end{equation}
This is suppressed by a $r$ factor with respect to the soft
masses.

Finally the classical A term vanishes. The $A$ term  picks a
quantum contribution
\begin{equation}
A_{an}\approx 3\gamma y \sqrt{\frac{1+2\alpha^2}{2}}m r
\end{equation}
which is suppressed by $\gamma r$ with respect to the  soft
masses.

We have thus characterized the mass spectrum of the superpartners.
The soft terms are all determined by the supersymmetry breaking
scale $F_{\mathcal G}$ and the value of the radion field $r$ where
$r$ is driven to zero according to the scalar potential.
 The conclusions are unmodified for the Randall-Sundrum
case $\alpha=0$.

Notice that the potential for the radion $r$ is now
\begin{equation}
V=\frac{3\kappa_4^2}{1+2\alpha^2}e^{-\frac{12\alpha^2}{1+2\alpha^2}\sigma}
|w_2|^2\sinh^{\frac{4-4\alpha^2}{1+2\alpha^2}}(r)
\end{equation}
driving the $r$ field to zero. As the field rolls down to its
minimum, the vacuum energy becomes  smaller and smaller. Notice
that for small $r$
\begin{equation}
V\approx \frac{3}{4(1+2\alpha^2)}\frac{\vert F_{\mathcal
G}\vert^2}{\kappa_4^2} r^{-2}
\end{equation}
Fine--tuning the cosmological constant for $\vert F_{\mathcal
G}\vert\approx 1$ TeV would require $r_{now}\approx 10^{45}$. To
remedy this problem, we now introduce an explicit supersymmetry
breaking step. This will allow us to obtain a very small vacuum energy,
so that the soft terms have the usual physical interpretation as masses and coupling constants in an almost flat space-time.

\section{Charged Branes \label{charged branes}}
The branes that we have considered so far are neutral branes. One
can also consider the case where a four--form lives in the bulk
and couples to the bulk scalar field. This induces a new
contribution to the potential for the moduli which turns out to be
of the same form as the potential obtained with $w_i\ne 0$
\cite{detuned}. The main difference here springs from the fact the
we introduce the brane charges in an explicitly supersymmetry
breaking form. It would be nice to embedd the charged case in a
fully supersymmetric framework. A possibility would consist in
using the Lagrange multiplier four--form of (\ref{boundary
action}), but then the added kinetic terms would have to be
supersymmetrised too.

Let us consider a four--form $C_{abcd}$ living in the bulk. We
define its field strength $F=dC$ which is a five--form dual to a
scalar field $* F$. The branes have charges $\pm Q$. The global
charge is zero as the fifth dimension is compact. We choose the
Lagrangian to be
\begin{widetext}
\begin{equation}
S_C=-\frac{1}{2} \int \frac{1}{U^2}F\wedge *F
-\frac{\sqrt{6(1-\alpha^2)}}{1+2\alpha^2}\frac{\kappa_5 Q}{\sqrt
k}\int_+ C +\frac{\sqrt{6(1-\alpha^2)}}{1+2\alpha^2}\frac{\kappa_5
Q}{\sqrt k}\int_-C \label{action}
\end{equation}
\end{widetext}
The prefactors are chosen for convenience. Notice that the
coupling constant for the four--form is proportional to $U$. One
can easily find the solutions of the equations of motion for $C$
\begin{equation}
*F=\frac{\sqrt{6(1-\alpha^2)}}{1+2\alpha^2}\frac{\kappa_5 Q}{\sqrt
k}\epsilon (z)  U^2
\end{equation}
where $\epsilon (z)$ is the odd function jumping from -1 to 1 at
the first brane and 1 to -1 at the second brane. Notice that the
boundary term depends on $\int F$ and vanishes altogether. The
action (\ref{action}) when evaluated yields a potential in the
Einstein frame which turns out to be
\begin{equation}
V= -\frac{3\kappa_4^2}{1+2\alpha^2}a(t_1)^{-12\alpha^2}\frac{Q^2
\rho^{4-4\alpha^2}-Q^2}{(1-\rho^{2+4\alpha^2})^2} \label{pot1}
\end{equation}
This is nothing but the potential for the moduli when substituting
$|w_{1,2}|^2 \to -Q^2$. In particular the total potential taking
into account both the tension detuning and the explicit
supersymmetry breaking by the charge $Q$ is now
\begin{equation}
V=
\frac{3\kappa_4^2}{1+2\alpha^2}a(t_1)^{-12\alpha^2}\frac{(|w_2|^2-Q^2)
\rho^{4-4\alpha^2}-(|w_1|^2-Q^2)}{(1-\rho^{2+4\alpha^2})^2}
\label{pot2}
\end{equation}
This potential has a much richer structure. We can distinguish
five cases. We will discuss the nature of the potential as a
function of $\rho=a(t_2)/a(t_1)$ (see figures \ref{figure
potentiels 1}, \ref{figure potentiels 2},  \ref{figure
potentiels 3}, 4, 5 and 6). \vskip .5 cm

\noindent {\it i) $\Big( |w_2| \leq |w_1|$ and $|Q| < |w_1|\Big)$
or $|w_2|<|w_1|=|Q|$} \vskip .5 cm

 The potential is negative and unbounded from
below. This case is not favoured phenomenologically. This
corresponds to figures \ref{figure potentiels 3}
 and \ref{figure potentiels 5}.

\vskip .5 cm

 \noindent {\it ii) $\vert Q\vert< \vert w_1\vert <
\vert w_2 \vert $}

\vskip .5 cm

 The potential admits a negative minimum leading to anti
de Sitter space. This is not a viable phenomenological case. Note
that this is the only case where radion stabilisation occurs. This
corresponds to figure \ref{figure potentiels 1} on the left hand
side. \vskip .5 cm

 \noindent {\it iii) $\vert Q\vert> \vert w_1\vert >
\vert w_2 \vert $}

\vskip .5 cm

 The potential is unbounded from below as $\rho \to
1$. Nevertheless it admits a local minimum at the origin
corresponding to a positive cosmological constant. The de Sitter
phase of the theory with $\rho=0$ is a metastable state protected
from the unstable branch of the potential by a local maximum.
Eventually the Universe would tunnel through this finite energy
barrier. The resulting phase would result in a big crunch
singularity due to the brane collision. This case corresponds to
figure \ref{figure potentiels 2}. \vskip .5
cm

\noindent {\it iv) $\Big(|w_1|\leq |w_2|$ and $|w_1|<Q \Big)$ or
$|w_2|>|w_1|=|Q|$}

\vskip .5 cm

 In that case the potential is bounded from below. Its
minimum is at the origin $\rho=0$. The corresponding cosmological
constant is positive and can be fine--tuned to match the present
value of the vacuum energy. This corresponds to figures
\ref{figure potentiels 4} and \ref{figure potentiels 6} on the
right hand sides.
 \vskip .5 cm

 \noindent {\it v) $|Q|=|w_1|=|w_2|$}

 \vskip .5 cm
The potential vanishes altogether, a situation reminiscent of a
BPS bound.

\section{Cosmological and Gravitational Consequences \label{grav and cosmo consequences}}

We will now investigate the cosmological consequences of the
previous model where supersymmetry is broken on the hidden brane.
The potential reads
\begin{widetext}
\begin{equation}
V=\frac{3\kappa_4^2}{1+2\alpha^2}e^{-\frac{12\alpha^2}{1+2\alpha^2}\sigma}
[Q^2 \cosh^{\frac{4-4\alpha^2}{1+2\alpha^2}}(r)
-(Q^2-|w_2|^2)\sinh^{\frac{4-4\alpha^2}{1+2\alpha^2}}(r)]
\end{equation}
\end{widetext}
Note that
$\rho=\frac{a(t_2)}{a(t_1)}=\tanh^{\frac{1}{1+2\alpha^2}}(r)$. The
field $r$ is driven towards its minimum at $r=0$. For small $r$
the potential reduces to an exponential potential for the sliding
field $\sigma$
\begin{equation}
V=\frac{3\kappa_4^2
\tilde Q^2(r)}{1+2\alpha^2}e^{-\frac{12\alpha^2}{1+2\alpha^2}\sigma}
\end{equation}
where we have defined $\tilde Q^2(r) \equiv Q^2
\cosh^{\frac{4-4\alpha^2}{1+2\alpha^2}}(r)
-(Q^2-|w_2|^2)\sinh^{\frac{4-4\alpha^2}{1+2\alpha^2}}(r) > 0$,
which reduces to the constant $Q^2$ for small enough $r$. This is
a typical example of a quintessence potential for $\sigma$. The
phenomenology of this type of potential is well--known. Let us
summarize some of its salient features.

The exponential potential admits an attractor with scale factor
\begin{equation}
a=a_0 t^{\frac{1+2\alpha^2}{3\alpha^2}}
\end{equation}
which is accelerating as soon as $\alpha <1$. Even in the presence
of cosmological matter such as cold dark matter, the above
attractor still exists and attracts the energy fraction carried by
the sliding field towards
\begin{equation}
\Omega_\sigma =1
\end{equation}
i.e. the Universe becomes scalar field dominated eventually. On
the attractor the equation of state of the Universe is constant
\begin{equation}
w=-1+ \frac{4\alpha^2}{1+2\alpha^2}
\end{equation}
We will see that $\alpha$ is constrained by solar system
experiments leading to a tight bound on the equation of state.

These results are equivalent to solving the full 5d equations of
motion with a positive detuning of the tension on the positive
tension brane. In 5d the motion of the positive tension brane is
at constant speed in conformal coordinates with speed $v$
\begin{equation}
\sqrt{1-v^2}= \frac{1}{1+ \kappa_5^2 \tilde Q^2}
\end{equation}
For small $\tilde Q$, the brane is non--relativistic and $v=\pm \sqrt 2
\kappa_5 \tilde Q$.

Of course, the Universe cannot be on the attractor now as one
expects $\Omega_\sigma \approx .7$. Hence one must fine--tune the
value of the sliding field now in such a way that
\begin{equation}
\tilde Q e^{-\frac{6\alpha^2}{1+2\alpha^2}\sigma_{now}}\approx\sqrt{\frac{1+2\alpha^2}{3\kappa_4^2}}\sqrt
{\Omega_\Lambda \rho_c}
\end{equation}
where $\Omega_\Lambda \approx 0.7$ and $\rho_c\approx 10^{-48}$
GeV$^4$. Numerically this leads to
\begin{equation}
\tilde Q e^{-\frac{6\alpha^2}{1+2\alpha^2}\sigma_{now}}\approx 10^{-6}
\hbox{GeV}^3
\end{equation}
This is nothing but the usual fine--tuning of the cosmological
constant. In particular, due to the smallness of the vacuum energy we can interpret the soft terms in the conventional way where the cosmological constant is set to zero. Notice that this is achieved thanks to the charge Q.

The fields $\sigma$ and $r$ are extremely light fields of masses
of the order of the Hubble rate $H_0\approx 10^{-42}$ GeV. These
extremely light fields may lead to large deviations from gravity
in solar system experiments. The solar system experiments are
mainly sensitive to the variation of the nucleon masses as a
function of the moduli. The bulk of the nucleon masses is given by
the QCD condensate $\Lambda_{QCD}$ expressed in the Einstein
frame. This condensate can be estimated from the pole of the
strong coupling constant at one loop and reads
\begin{equation}
\Lambda_{QCD}=A(\sigma,r) e^{-\frac{2\pi}{b_3}{\mathcal R}e\ f_3(\sigma,r,{\cal I}m
T_1)} \Lambda_0
\end{equation}
where $f_3$ is the strong coupling function in the Lagrangian and
$b_3$ is the QCD renormalisation group coefficient (the 3 index
signaling the SU(3) gauge group). The overall factor $A(\sigma,
r)$ accounts for the change of frame between the brane frame and
the Einstein frame. The scale $e^{-\frac{2\pi}{b_3}{\mathcal R}e\
f_3(\Lambda_0)}\Lambda_0$ is the renormalisation group invariant
QCD scale in the brane frame where ordinary matter is minimally
coupled to the brane metric. Explicitly we find that
\begin{equation}
A(\sigma ,r)= e^{-\frac{2\alpha^2}{1+2\alpha^2}\sigma}
\cosh^{\frac{1}{1+2\alpha^2}}(r)
\end{equation}
The solar system experiments give tight bounds on the Eddington
parameter $\vert \gamma_{ED}-1\vert \le 10^{-5}$ \cite{testsGR}.
The Eddington parameter is related to $\gamma_{ED}-1\approx -2
\theta$ where $\theta$ depends the coupling constants
\begin{equation}
\alpha_{r}=\frac{\partial \ln \Lambda_{QCD}}{\partial
r},\alpha_{\sigma}=\frac{\partial \ln \Lambda_{QCD}}{\partial
\sigma},\alpha_{{\cal I}m T_1}=\frac{\partial \ln \Lambda_{QCD}}{\partial
{\cal I}m T_1}
\end{equation}
and the normalisation of the fields $\gamma_{ij}$, with inverse
matrix $\gamma^{ij}$. It is defined by

\begin{equation}
\theta= \gamma^{ij}\alpha_i\alpha_j=\gamma^{rr}\alpha_r^2
+\gamma^{\sigma \sigma}\alpha_\sigma^2 +\frac{1}{2}K^{T_1\bar
T_1}\alpha_{{\cal I}m T_1}^2 .
\end{equation}

This constraint is valid for massless scalar fields, in the sense
as their inverse mass is larger than solar system scales. This is
the case for the moduli we consider whose inverse mass is
typically of the size of the Hubble radius. Explicitly this leads
to
\begin{widetext}
\begin{eqnarray}
\theta&=&\frac{1}{3}\frac{\alpha^2}{1+2\alpha^2}\Big(1+\frac{\pi}{b_3}{\mathcal R}ef_3\frac{\beta_3}{\alpha^2}(\frac{a(t_1)}{a(T_1)})^{4\alpha^2}\Big)^2+
\frac{\tanh^2
r}{6(1+2\alpha^2)}\Big(1-\frac{2\pi}{b_3}{\mathcal R}ef_3\ \beta_3(\frac{a(t_1)}{a(T_1)})^{4\alpha^2}\Big)^2
\nonumber \\
&&+\frac{1}{12(1+2\alpha^2)}\Big(1+2\alpha^2 \tanh^2(r)\Big)
\frac{4\pi^2}{b_3^2} \Big( {\mathcal R}ef_3\frac{\beta_3}{\alpha} (\frac{a(t_1)}{a(T_1)})^{4\alpha^2}\Big)^2 . \label{Eddington}
\end{eqnarray}
\end{widetext}
Note that $\beta^i_a \equiv \frac{\partial \ln
{\mathcal R}e f^i_a}{\partial \ln a(T_i)}$ is now defined from the
\emph{real} gauge coupling ${\mathcal R}e f^i_a$ and not any
longer from its complex version.

Imposing the experimental bounds leads to
\begin{equation}
\alpha <10^{-2},\ r_{now}\le 10^{-2}
\end{equation}
together with the fact that $\frac{{\mathcal R}e
f_3}{\alpha}\vert\beta_3\vert=\frac{1}{\alpha}|\frac{\partial
{\mathcal R}e f_3}{\partial ln a(T_1)}|$ cannot be too large
compared to one. Considering the part of moduli-dependence due to
the coupling of the gauge kinetic terms with the brane value of
the bulk scalar field $\phi(T_1)=-4\alpha\ \ln(a(T_1))$, one has
$\frac{{\mathcal R}e f_3}{\alpha} |\beta_3|=4|\frac{\partial
{\mathcal R}e f_3}{\partial \phi(T_1)}|$ which has to be small
compared to one.

The QCD gauge coupling $f_3$ is of order one and the factor
$|\frac{a(t_1)}{a(T_1)}|$ is always smaller than one. The bound on
$r_{now}$ can be fulfilled as the value $r=0$ is a minimum of the
potential. Choosing $\alpha$ small enough, we find that the
scalar--tensor theory is attracted towards general relativity.
Notice that this leads to a very tight constraint on the equation
of state of matter on the attractor $w_\sigma\le -0.96$ hardly
distinguishable from a pure cosmological constant.

Now let us come back to the supergravity case with a vector
multiplet where $\alpha^2=1/12,1/3$. As can be seen the sliding
field $\sigma$ leads to large deviations from ordinary gravity
which are not compatible with experiments. The
$\gamma^{\sigma\sigma}\alpha_{\sigma}^2$ contribution to the
$\theta$ parameter is indeed the one proportional to $\alpha^2$
and thus too large in the vector multiplet case. Of course this is
the usual stabilisation problem as already appearing for the
dilaton. One possibility indeed would be to stabilise the $\sigma$
field. This would require to find a potential with an appropriate
minimum. This is a notoriously difficult problem. Let us conclude
with the pure Randall-Sundrum case $\alpha=0$. The quintessence
field $\sigma$ disappears so that the gravitational constraints
are now satisfied ; the vacuum energy  is adjusted to its present
value through the $\tilde Q$ parameter. This is the fine--tuning
of the cosmological constant.

Hence in the Randall-Sundrum case with supersymmetric matter on
the positive tension brane and detuning of the negative tension
brane, the gravitational constraints are satisfied for far--away
branes. In that case, the soft terms are determined by the anomaly
mediation breaking term $F_{\cal G}$ and the value of the radion
now. The radion is driven to zero implying that gravity is
retrieved and the soft masses are very large compared to the
gaugino masses and the $\mu$ term. For small values of $r$ this
would lead to a naturalness  problem for the electro--weak
breaking. The detailed analysis of the phenomenology of these
models is left for future work.

\section{Conclusion}

We have analysed the supersymmetry breaking of 5d supergravity
models with boundary branes in the presence of one vector
multiplet and the universal hypermultiplet. As a particular case,
this includes the supersymmetric Randall--Sundrum model. We have
discussed the soft breaking terms and the moduli potential when
supersymmetry is broken on both branes. We have focused on the
case where matter is present on the positive tension brane and
supersymmetry is broken on the negative tension brane. We have
seen that the spectrum is determined both by the anomaly mediation
$F_{\cal G}$ term and the normalised radion field $r$. In
particular the radion $r$ is driven to zero where general
relativity is retrieved. In that case the soft masses are much
larger than the gaugino masses leading to a possible naturalness
problem.

We have also discussed the case with the universal hypermultiplet
and in particular brane race--track models. These models are very
intricate and would deserve further study. In particular one may
hope that the moduli may become stabilised at appropriate values.

We have seen that the supergravity models with a vector multiplet
are disfavoured as leading to strong deviations from general
relativity. One way of modifying this result would be for the
moduli to become chameleon fields whereby their masses depend on
the environment \cite{chameleon}. In the solar system, this could
be enough to evade the gravitational constraints and enlarge the
class of phenomenologically relevant models. This is left for
future work.
\\

\leftline{\Large \bf Acknowledgements}
\

We would like to thank U. Ellwanger and Z. Lalak for useful
comments and discussions.

\appendix

\section{$AdS_4$ Supersymmetry and Radion stabilisation}
Let us consider the Randall-Sundrum brane system with detuned
tensions $\lambda_{1,2}=\pm \delta_{1,2}6k$ where
\begin{equation}
\delta_{1,2}=1-\kappa_5^2|w_{1,2}|^2
\end{equation}
where we assume that the detuning is small
\begin{equation}
\kappa_5^2|w_{1,2}|^2 \ll 1 .
\end{equation}
The solution of the 5d equations of motion including the brane
boundary condition leads to a fixed value of the interbrane
distance
\begin{equation}
kt=\frac{1}{2}\ln\Big(\frac{(\lambda+\lambda_1)(\lambda-\lambda_2)}{(\lambda+\lambda_2)(\lambda-\lambda_1)}\Big)
\end{equation}
where $\lambda=6k$ is the tuned RS tension. For small detuning
this gives
\begin{equation}
kt=\ln|\frac{w_2}{w_1}| .
\end{equation}
Notice that stabilisation of the radion can only be achieved when
$|w_2|>|w_1|$. In that case, the configuration is known to respect
$AdS_4$ supersymmetry. We will compare this 5d analysis to the
effective action approach.

In 4d at low energy, the K\"ahler potential
\begin{equation}
K=-3 \ln (e^{-k(T_1+\bar T_1)}-e^{-k(T_2+\bar T_2)})
\end{equation}
complemented with the superpotential
\begin{equation}
W=w_1 e^{-3kT_1} + w_2 e^{-3k T_2}
\end{equation}
is equivalent to
\begin{eqnarray}
K&=&-3 \ln (1-e^{-k(T+\bar T)}) \nonumber \\
W&=&w_1+w_2e^{-3kT}
\end{eqnarray}
up to a K\"ahler transformation, with $T=T_2-T_1$. The scalar
potential admits a minimum when $|\frac{w_2}{w_1}|>1$ where
\begin{equation}
kt=\ln|\frac{w_2}{w_1}|
\end{equation}
coinciding with the 5d result. At the minimum the $F^T$ term reads
\begin{equation}
F^T=-\frac{W}{|W|}\frac{\kappa_4^2}{k} \bar
w_1(1-|\frac{w_1}{w_2}|^2)^{-1/2}(1+e^{3 i k {\cal I}m T-i\eta})
\end{equation}
where $\eta$ is the phase of $\frac{w_2}{w_1}$. This vanishes for
\begin{equation}
3k{\cal I}m T=\eta + \pi (mod\ 2\pi)
\end{equation}
leading to a configuration with a negative potential energy
\begin{equation}
V=-3\frac{m_{3/2}^2}{\kappa_4^2}
\end{equation}
and a non zero gravitino mass. The solution of Einstein's
equations is then $AdS_4$ which breaks Poincar\'e invariance.
\\

\section{Soft terms in the Randall-Sundrum case $\alpha=0$}
In the Randall-Sundrum limit, only one chiral superfield out of
the $T_{1,2}$ remains physical. We start from
\begin{eqnarray}
K&=&-3 ln(1-e^{-k(T+\bar T)}) \nonumber \\
W&=&w_1+w_2 e^{-3kT}
\end{eqnarray}
where $T=T_2-T_1$ is the radion superfield, and we will write
$T=t+i{\cal I}m T$. As can be checked the results will be the same
as those obtained from our general calculation by taking the limit
$\alpha=0$. We write the soft Lagrangian for the canonically
normalised complex matter scalars $\tilde{s}_i$ on the \emph{positive}
tension brane as
\begin{equation}
{\cal L}_{soft}= -m^2_{i\bar j} \tilde{s}_i \overline{\tilde{s}_j}
- (\frac{1}{2}B_{ij}\tilde{s}_i \tilde{s}_j + hc) - (\frac{1}{6}
A_{ijk}\tilde{s}_i \tilde{s}_j \tilde{s}_k + hc) .
\end{equation}

The radion potential
\begin{equation}
V=-3\kappa_4^2\frac{|w_1|^2-e^{-4kt}|w_2|^2}{(1-e^{2kt})^2}
\end{equation}
does not depend on ${\cal I}m T$. The gravitino mass is
\begin{equation}
m_{3/2}=\kappa_4^2\frac{|w_1+w_2 e^{-3kT}|}{(1-e^{-2kt})^{3/2}} .
\end{equation}
The radion F--term is
\begin{equation}
F^T=-\frac{\bar W}{|W|}\frac{\kappa_4^2}{k}\frac{\bar w_1+\bar w_2
e^{-kt+3ik {\cal I}m T}}{(1-e^{-2kt})^{1/2}} .
\end{equation}
The effective normalised superpotential is given by
\begin{eqnarray}
\mu_{ij}^{eff}&=&\frac{\bar W}{|W|}\frac{1}{(1-e^{-2kt})^{1/2}}(\mu_{ij}+\kappa_4^2 \lambda_{ij} w_1)
\nonumber \\
\lambda^{eff}_{ijk}&=&\frac{\bar W}{|W|}\lambda_{ijk} .
\end{eqnarray}
The normalised soft mass is
\begin{equation}
m^2_{i\bar j}=\delta_{i \bar j}\frac{2}{3}\kappa_4^2
V=-2\delta_{i\bar
j}\kappa_4^4\frac{|w_1|^2-e^{-4kt}|w_2|^2}{(1-e^{2kt})^2}
\end{equation}
and the normalised soft bilinear term is
\begin{eqnarray}
B_{ij}&=& - \frac{\kappa_4^2 \bar w_1 }{1-e^{-2kt}}\mu_{ij}
+\kappa_4^4 \lambda_{ij}\frac{1}{(1-e^{-2kt})^2}
\nonumber \\
&&\Big(-|w_1|^2(1+e^{-2kt})
+2|w_2|^2e^{-4kt}\Big)
.
\end{eqnarray}
The normalised soft trilinear term is vanishing :
\begin{equation}
A_{ijk}= 0 .
\end{equation}

\end{document}